\documentclass[twocolumn]{aastex62}
\usepackage{color,soul}
\usepackage{comment}
\usepackage{amsmath}
\usepackage{tablefootnote}

\graphicspath{{./}{figures/}}

\defcitealias{Seitenzahl13b}{S13}
\defcitealias{Leung18}{L18}
\defcitealias{Fink14}{F14}
\defcitealias{Leung19}{L19}
\defcitealias{Shen18a}{S18}
\defcitealias{Bravo19}{B19}

\submitjournal{ApJ}

\shorttitle{Manganese as a Probe of Type Ia Progenitors}
\shortauthors{de los Reyes et al.}

\begin{document}

\title{Manganese Indicates a Transition from Sub- to Near-Chandrasekhar Type Ia Supernovae in Dwarf Galaxies\footnote{The data presented herein were obtained at the W. M. Keck Observatory, which is operated as a scientific partnership among the California Institute of Technology, the University of California and the National Aeronautics and Space Administration. The Observatory was made possible by the generous financial support of the W. M. Keck Foundation.}}

\correspondingauthor{Mithi A. C. de los Reyes}
\email{mdelosre@caltech.edu}

\author[0000-0002-4739-046X]{Mithi A. C. de los Reyes}
\affiliation{Department of Astronomy, California Institute of Technology \\
1200 E. California Blvd., MC 249-17 \\
Pasadena, CA 91125, USA }

\author{Evan N. Kirby}
\affiliation{Department of Astronomy, California Institute of Technology \\
1200 E. California Blvd., MC 249-17 \\
Pasadena, CA 91125, USA }

\author{Ivo R. Seitenzahl}
\affiliation{School of Science, University of New South Wales, Australian Defence Force Academy \\
Canberra, ACT 2600, Australia}

\author{Ken J. Shen}
\affiliation{Department of Astronomy and Theoretical Astrophysics Center, University of Califiornia \\
Berkeley, CA 94720, USA }

\begin{abstract}

Manganese (Mn) abundances are sensitive probes of the progenitors of Type Ia supernovae (SNe).
In this work, we present a catalog of manganese abundances in dwarf spheroidal satellites of the Milky Way, measured using medium-resolution spectroscopy.
Using a simple chemical evolution model, we infer the manganese yield of Type Ia SNe in the Sculptor dwarf spheroidal galaxy (dSph) and compare to theoretical yields.
The sub-solar yield from Type Ia SNe ($\mathrm{[Mn/Fe]}_{\mathrm{Ia}}=-0.30_{-0.03}^{+0.03}$ at $\mathrm{[Fe/H]}=-1.5$~dex, with negligible dependence on metallicity) implies that sub-Chandrasekhar-mass (sub-$M_{\mathrm{Ch}}$) white dwarf progenitors are the dominant channel of Type Ia SNe at early times in this galaxy, although some fraction ($\gtrsim20\%$) of $M_{\mathrm{Ch}}$ Type Ia or Type Iax SNe are still needed to produce the observed yield.
First-order corrections for deviations from local thermodynamic equilibrium increase the inferred $\mathrm{[Mn/Fe]}_{\mathrm{Ia}}$ by as much as $\sim0.3$~dex.
However, our results also suggest that the nucleosynthetic source of Type Ia supernovae may depend on environment.
In particular, we find that dSph galaxies with extended star formation histories (Leo I, Fornax dSphs) appear to have higher [Mn/Fe] at a given metallicity than galaxies with early bursts of star formation (Sculptor dSph), suggesting that $M_{\mathrm{Ch}}$ progenitors may become the dominant channel of Type Ia SNe at later times in a galaxy's chemical evolution.
\end{abstract}


\section{Introduction} 
\label{sec:intro}
Type Ia supernovae (Ia SNe) have long been understood to be the thermonuclear explosions of white dwarfs.
Their ability to be empirically normalized to the same peak luminosity \citep[e.g.,][]{Phillips93} has made them indispensible astrophysical tools as ``standardizable candles'' for measuring cosmological distances.
Indeed, Type Ia SNe were used in the Nobel Prize-winning discovery of the accelerating expansion of the Universe \citep{Riess98,Perlmutter99}.

However, the fundamental physics governing Type Ia SNe---particularly the actual explosion mechanism---are still poorly constrained.
The traditional paradigm of Type Ia SNe suggests that a thermonuclear supernova occurs when a single white dwarf (WD) accretes material from a non-degenerate companion star and undergoes runaway thermonuclear burning near the Chandrasekhar mass of $M_{\mathrm{Ch}}\approx1.4~M_{\odot}$.

Several problems persist with this paradigm.
Simulated detonations of a $M_{\mathrm{Ch}}$ white dwarf tend to underproduce intermediate-mass elements (IMEs) such as silicon and sulfur that dominate observed Type Ia SNe light curves \citep[e.g.,][]{Arnett71}.
Near-Chandrasekhar mass white dwarfs also appear to be rare
\citep[e.g.,][]{Tremblay16,Giammichele12}, and increasing the mass of an accreting white dwarf can be challenging \citep[e.g.,][]{Shen07,Maoz14}.
Finally, this physical mechanism requires accretion from a companion star, but multiple nearby Type Ia SNe have been observed without companions.
For example, radio and X-ray data from SN2011fe in M101 strongly disfavor the existence of a non-degenerate companion star \citep{Margutti12, PerezTorres14}.

Various models have attempted to resolve some of these discrepancies, largely by tweaking the assumptions of a \emph{prompt detonation} of a \emph{single} white dwarf at the \emph{Chandrasekhar} limit.
For example, if a $M_{\mathrm{Ch}}$ white dwarf is allowed to expand before detonation instead of promptly detonating, the expansion will produce low-density regions.
These regions then provide ideal conditions for the nucleosynthesis of the missing IMEs \citep{Seitenzahl17}.
One way to achieve this scenario is by prolonging the explosion with the so-called ``deflagration-to-detonation transition (DDT)'' \citep{Khokhlov91}.

Alternatively, the rarity of $>1$~M$_{\odot}$ white dwarfs suggests that many, if not most, Type Ia SNe are produced by the explosions of lower-mass WDs.
One of the most favored models for exploding a single sub-$M_{\mathrm{Ch}}$ white dwarf is the ``double detonation'' model, in which the WD accretes helium from a He-rich companion, such as a non-degenerate He-star.
The helium shell may detonate when it becomes massive enough, sending shocks through the white dwarf that explode it \citep{Nomoto82b,Woosley86,Livne90}.
This model can successfully reproduce most observations of typical Type Ia SNe, including the nucleosynthesis of IMEs \citep[e.g.,][]{Woosley11}.

Finally, a sub-$M_{\mathrm{Ch}}$ white dwarf can also explode if it has a second white dwarf companion.
This ``double degenerate'' channel may account not only for the rarity of massive white dwarfs and the expected nucleosynthesis of IMEs, but also for the missing companion stars near some observed Type Ia SNe.
Physically, a double degenerate explosion may be similar to the double detonation model described above, in which the primary WD accretes from a secondary He WD \citep[e.g.,][]{Shen18a}.
This model has been invoked to explain the discovery of hypervelocity white dwarfs, which are thought to be surviving donor companions of these ``dynamically driven double-degenerate double-detonation'' explosions \citep{Shen18b}.
Alternatively, binary sub-$M_{\mathrm{Ch}}$ white dwarfs can merge, form a super-$M_{\mathrm{Ch}}$ remnant, and undergo a deflagration-to-detonation transition \citep{Iben84,Webbink84}.

The abundances of elements produced by Type Ia SNe can be used to distinguish between these physical models.
While these abundances can be measured directly in spectra of supernovae or supernova remnants, these direct observations are inherently limited by the rarity of Type Ia SNe, and many abundances are difficult to determine from direct spectroscopy.
The focus of this paper is to instead indirectly infer nucleosynthetic yields from ancient stars, because the abundances of these stars are linked to the chemical evolution of a galaxy.

\subsection{Measuring nucleosynthesis with dwarf galaxies}
\label{sec:dwarfs}
The chemical evolution of a galaxy is largely driven by enrichment from supernovae.
Core-collapse supernovae are driven by the deaths of the most massive stars in a galaxy, which begin very early in a galaxy's history.
Type Ia supernovae, on the other hand, can only begin to explode much later, after lower-mass stars die and create white dwarfs.

Both Type Ia and core-collapse SNe produce iron. 
Throughout the lifetime of a galaxy, SNe will therefore produce an increase in the overall abundance of iron, [Fe/H]\footnote{Throughout this paper, we use bracket abundances referenced to solar (e.g., [Fe/H] = $\log_{10}(n_{\mathrm{Fe}}/n_{\mathrm{H}})_{\ast}-\log_{10}(n_{\mathrm{Fe}}/n_{\mathrm{H}})_{\odot}$), where $n_{\mathrm{X}}$ is the atomic number density of X. Solar abundances are adopted from \citet{Asplund09}.}.
However, because Type Ia and core-collapse SNe have different nucleosynthetic products, the abundance of other elements relative to iron will change once Type Ia SNe begin to explode.
In particular, since dwarf spheroidal (dSph) galaxies have low star formation rates, their chemical evolution is dominated at late times by Type Ia SNe rather than by core-collapse SNe.
As [Fe/H] increases over time, the relative abundance of an element relative to iron will approach the Type Ia yield.
The yields of various elements can then be used to infer properties of Type Ia SNe alone \citep{McWilliam18}.

Furthermore, the abundance contributions specifically from Type Ia SNe ($f_{\mathrm{Ia}}$) can be computed using the well-constrained theoretical yields of various elements from core-collapse SNe.
Once the Type Ia SNe yields are disentangled from core-collapse SNe yields, measurements of different elemental abundances can be used to infer properties of Type Ia SNe alone.
\citet{Kirby19} originally performed this analysis for several iron-peak elements (Cr, Co, Ni), fitting a simple chemical decomposition model to determine $f_{\mathrm{Ia}}$ and measure the absolute Type Ia yields of these elements.
These yields suggested that sub-$M_{\mathrm{Ch}}$ white dwarfs are the dominant progenitors of Type Ia SNe in dwarf galaxies at early times. 
\citet{Kirby19} also found that galaxies with star formation lasting for several Gyr have higher [Ni/Fe] abundances than galaxies with an early burst of star formation, potentially indicating that the dominant Type Ia supernova channel depends on star formation history.

\subsection{Manganese}
\label{sec:mn}
In this work, we aim to extend the analysis of \citet{Kirby19} to manganese (Mn), which is a particularly sensitive probe of the physics of Type Ia SNe \citep{Seitenzahl09,Seitenzahl13,Seitenzahl15}.
Like the other iron-peak elements, the production of Mn is dominated by Type Ia rather than core-collapse SNe.
Furthermore, the only stable isotope of manganese, $^{55}$Mn, is produced via nucleosynthetic pathways that are strongly dependent on the density of the progenitor white dwarf.

Nearly all $^{55}$Mn is produced as its radioactive parent nucleus $^{55}$Co, which can be produced in low entropy (``normal'') freeze-out from nuclear statistical equilibrium at densities $\rho\gtrsim 2\times10^{8}$~g~cm$^{-3}$ \citep{Seitenzahl17}. 
Higher yields of $^{55}$Co and therefore $^{55}$Mn can be achieved if silicon does not completely burn, while lower yields can be achieved at high entropy and low density, where the presence of protons during ``alpha-rich'' freeze-out ultimately destroys $^{55}$Co via the reaction $^{55}$Co(p,$\gamma$)$^{56}$Ni \citep{Seitenzahl13}.
In white dwarfs well below $M_{\mathrm{Ch}}$, $^{55}$Co is generally produced at densities below nuclear statistical equilibrium, producing lower yields of $^{55}$Mn.

In other words, stable Mn is more likely to be synthesized in near-$M_{\mathrm{Ch}}$ white dwarfs than in sub-$M_{\mathrm{Ch}}$ progenitors.
The observed yield of Mn from Type Ia SNe is therefore physically significant---higher yields suggest $M_{\mathrm{Ch}}$ explosions, while lower yields may indicate sub-$M_{\mathrm{Ch}}$ models.
To that end, there is significant interest in measuring stellar manganese abundances.

Previous works have presented conflicting results of Mn measurements in nearby dSphs.
\citet{North12} compiled literature Mn abundances and used high-resolution spectroscopy to measure additional Mn abundances for stars in Sculptor ($N=50$) and Fornax ($N=60$) dSphs.
They concluded that the Mn abundances imply sub-solar [Mn/Fe] ratios, and that the specific trend of [Mn/Fe] vs [Fe/H] implies a metallicity-dependent Mn yield from Type Ia SNe.
However, the \citet{North12} measurements used high-resolution spectroscopy and were largely limited to higher-metallicity stars ($\mathrm{[Fe/H]}\gtrsim-1.75$), making it difficult to precisely constrain the trend of [Mn/Fe] over a large range of [Fe/H].

On the other hand, \citet{Kobayashi15} used a different sample to suggest that high Mn abundances point to dense Type Ia SNe, and that a special class of near-$M_{\mathrm{Ch}}$ ``Type Iax'' SNe are needed to produce enough Mn to match observations.
\citet{Cescutti17} made a similar argument for a combination of ``normal'' and ``Iax'' SNe using Mn abundances for $N=20$ stars in the dSph Ursa Minor.
In both studies, the observations are too incomplete to draw any significant conclusions.

In this paper, we increase the sample size and parameter space of these literature Mn abundances by using medium-resolution spectra to extend to fainter and more metal-poor stars in dSph galaxies.
We then use these measurements to distinguish between different Type Ia SNe models.
Our observations are described in Section~\ref{sec:observations}.
In Section~\ref{sec:abundances}, we describe our pipeline for measuring Mn abundances, validate our measurement technique using globular clusters, and present Mn abundances for stars in classical dSph galaxies.
We use a simple chemical evolution model to infer Mn yields from Type Ia SNe in Section~\ref{sec:typeia} before discussing the implications for Type Ia SN physics in Section~\ref{sec:implications}.
Finally, we summarize our conclusions in Section~\ref{sec:summary}.

\section{Observations} \label{sec:observations}

Unlike literature catalogs, which generally use high-resolution spectra to measure abundances, this work aims to use medium-resolution spectra to measure Mn abundances.
Medium-resolution spectroscopy was performed using the DEep Imaging Multi-Object Spectrograph \citep[DEIMOS;][]{Faber03} on the Keck II telescope.
Spectra were obtained for red giant branch (RGB) stars in several globular clusters and classical dSphs.
Table~\ref{tab:targets} lists the observations of the globular clusters and dSphs used in this work.

Our target selection prioritizes globular clusters and dSphs previously observed with the red 1200G grating on DEIMOS.
We used a combination of old and newly designed slitmasks. \citet{Kirby09,Kirby10,Kirby16} presented 1200G observations of bscl5, bfor7, n5024b (previously called ng5024), 7078l1, and 7089l1.  The masks LeoIb, CVnIa, and bumia are very similar to other masks observed by \citet{Kirby10}, but previous observations allowed us to determine membership for some stars.  We designed the new masks to have fewer non-members and more confirmed members.  We did the same for UMaIIb, where \citet{Simon07} observed the previous slitmasks for Ursa Major~II\@.

The previous references describe the membership selection, which we adopt here.  In general, members were selected to have radial velocities within 3$\sigma_v$ of the mean velocity.  They were also required to have colors and magnitudes consistent with the red giant branches of their respective galaxies.

In this work we used the 1200B grating, which was commissioned in September 2017.  The grating has a groove spacing of 1200~mm$^{-1}$ and a blaze wavelength of 4500~\AA\@.  It provides a dispersion of 0.34~\AA~pixel$^{-1}$ for first-order light.  The FWHM of the line spread function is about 1.1~\AA\@.  The corresponding resolving power at 5000~\AA\ is $R = \lambda/\Delta\lambda = 4500$.  In contrast to DEIMOS's previous complement of gratings, 1200B provides higher resolution than 900ZD and higher throughput at $\lambda < 6000$~\AA\ than 1200G\@.

We used a central wavelength of 5200~\AA, which provided an approximate spectral range of 3900--6500~\AA, but the exact spectral range for each slit depended on the location of the slit on the slitmask.  The variation in the starting and ending wavelengths was as much as 250~\AA\@.  The GG400 order-blocking filter eliminated light bluer than 4000~\AA\ so that second-order light did not contaminate our spectra.

\begin{deluxetable*}{lllrlclccc}
\tablecolumns{10} 
\tablecaption{ Spectroscopic targets. \label{tab:targets}} 
\tablehead{ 
\colhead{Object} & \colhead{RA} &  \colhead{Dec} & \colhead{Dist.} & \colhead{Slitmask} & \colhead{$N_{\mathrm{stars}}$} & \colhead{Date} & \colhead{Airmass} & \colhead{Seeing} & \colhead{Exposures} \\
\colhead{} & \colhead{(J2000)} & \colhead{(J2000)} & \colhead{(kpc)} & \colhead{} & \colhead{} & \colhead{} & \colhead{} & \colhead{('')} & \colhead{(s)}
}
\startdata
\multicolumn{10}{c}{Globular clusters}\\[0.5em]
\tableline
M53 (NGC 5024) & 13$^{\mathrm{h}}$12$^{\mathrm{m}}$55$^{\mathrm{s}}$ & +18$^{\circ}$09'59'' & 17.9 & n5024b & 182 & 2019 Mar 10 & 1.0 & 1.6 & 5$\times$1200 \\
               &                                                     &                     &     &               &     & 2019 Mar 11 & 1.0 & 0.9 & 1$\times$404 \\[0.5em]
M15 (NGC 7078) & 21$^{\mathrm{h}}$29$^{\mathrm{m}}$49$^{\mathrm{s}}$ & +12$^{\circ}$10'20'' & 10.4 & 7078l1 & 175 & 2017 Sep 15 & 1.1 & 0.6 & 13$\times$1200 \\[0.5em]
M2 (NGC 7089) & 21$^{\mathrm{h}}$33$^{\mathrm{m}}$15$^{\mathrm{s}}$ & $-$00$^{\circ}$48'36'' & 11.5 & 7089l1 & 157 & 2017 Oct 3 & 1.1 & \nodata & 3$\times$1200, 1$\times$1800 \\[0.5em]
\tableline
\multicolumn{10}{c}{dSphs}\\[0.5em]
\tableline
Sculptor & 00$^{\mathrm{h}}$59$^{\mathrm{m}}$57$^{\mathrm{s}}$ & $-$33$^{\circ}$41'45'' & 86 & bscl5 & 97 & 2018 Aug 14 & 1.8 & 0.8 & 3$\times$1500 \\
         &                                                     &                      &    &              &     & 2018 Sep 10 & 1.8 & 0.7 & 3$\times$1800, 1$\times$860 \\
         &                                                     &                      &    &              &     & 2018 Sep 11 & 1.8 & 0.8 & 2$\times$1800 \\[0.5em]
Fornax & 02$^{\mathrm{h}}$39$^{\mathrm{m}}$49$^{\mathrm{s}}$ & $-$34$^{\circ}$30'35'' & 147 & bfor7 & 154 & 2018 Aug 14 & 1.8 & 0.9 & 2$\times$1560, 1$\times$1440 \\
         &                                                     &                      &    &              &     & 2018 Sep 10 & 1.8 & 0.7 & 2$\times$1320, 2$\times$1620 \\
         &                                                     &                      &    &              &     & 2018 Sep 11 & 2.0 & 0.8 & 2$\times$1980 \\[0.5em]
Ursa Major II & 08$^{\mathrm{h}}$52$^{\mathrm{m}}$48$^{\mathrm{s}}$ & +63$^{\circ}$05'54'' &  32 & UMaIIb &  21 & 2019 Feb 6 & 1.5 & \nodata & 3$\times$1740 \\[0.5em]
Leo I & 10$^{\mathrm{h}}$08$^{\mathrm{m}}$29$^{\mathrm{s}}$ & +12$^{\circ}$18'56'' & 254 & LeoIb & 137 & 2018 Mar 19 & 1.3 & 0.8 & 2$\times$1620, 1$\times$1560 \\
      &                                                     &                      &    &              &     & 2019 Feb 6  & 1.1 & \nodata & 2$\times$1860, 1$\times$1920 \\
      &                                                     &                      &    &              &     & 2019 Mar 12 & 1.2 & 0.8 & 3$\times$1800, 2$\times$1500 \\[0.5em]
Canes Venatici I & 13$^{\mathrm{h}}$28$^{\mathrm{m}}$03$^{\mathrm{s}}$ & +33$^{\circ}$32'44'' & 218 & CVnIa & 125 & 2018 Mar 19 & 1.1 & 0.7 & 3$\times$1680, 2$\times$1860 \\
                 &                                                     &                      &    &              &     & 2018 May 20 & 1.0 & 1.0 & 1$\times$1200, 2$\times$906 \\
                 &                                                     &                      &    &              &     & 2019 Mar 12 & 1.2 & 0.8 & 6$\times$1800 \\[0.5em]
Ursa Minor & 15$^{\mathrm{h}}$08$^{\mathrm{m}}$32$^{\mathrm{s}}$ & +67$^{\circ}$11'03'' &  76 & bumia & 135 & 2019 Mar 12 & 1.5 & 1.4 & 4$\times$1800, 2$\times$2100 \\[0.5em]
\enddata
\end{deluxetable*}

Table~\ref{tab:targets} details the observations for each field.  We observed one slitmask per globular cluster or dwarf galaxy.  The coordinates indicate the center of the slitmasks, not necessarily the centers of the stellar systems.  The distances are taken from \citet{Harris96} for globular clusters and \citet{McConnachie12} for dwarf galaxies.  The number of stars represents the total number of slits, including both members and non-members of the corresponding stellar systems.  We also report the average airmass and seeing (where available) for the observations.

All observations were reduced using a version of the \texttt{spec2d} pipeline \citep{Newman13, Cooper12}.  The pipeline traces the edges of the slits with the help of a spectrally dispersed image of a quartz continuum lamp.  The same spectral frame provides for flat fielding.  We used separate exposures of Ne, Ar, Kr, Xe, and Hg arc lamps for wavelength calibration.  We identified arc lines with the help of the NIST atomic spectra database \citep{Kramida14}.  After flat fielding and wavelength calibration, the \texttt{spec2d} pipeline performs sky subtraction in 2-D and then extracts the spectra into 1-D\@. We modified \texttt{spec2d} in several ways to improve the reliability of the wavelength solution for the 1200B grating.  Most notably, we changed one of the subroutines that determined whether an arc line should be included in the wavelength calibration so that usable arc lines were not discarded erroneously.

DEIMOS uses active flexure compensation to keep the data frames aligned within $\sim 0.1$~pixel in both the spatial and spectral directions.  The flexure compensation allowed us to stack the 2-D images taken within the same week.  However, the compensation becomes unreliable beyond about a week.  Over longer timescales, the heliocentric velocity correction varies too much to stack images.  Therefore, we reduced images taken within the same week into 1-D spectra.  For slitmasks observed over multiple weeks, we coadded the 1-D spectra after correcting for the change in the heliocentric reference frame.

\section{Abundance Measurements} \label{sec:abundances}

\subsection{Description of Pipeline}
\label{sec:smaug}
In this section, we describe the analysis pipeline used to obtain Mn abundances from the reduced and corrected spectra.
Broadly speaking, this pipeline fits synthetic spectra with variable Mn abundances to an observed spectrum and uses least-squares fitting to determine the Mn abundance.

\subsubsection{Inputs}
\label{sec:inputs}

The main inputs to this pipeline are a line list---a list of atomic and molecular lines in the spectral regions of interest---and estimates of stellar parameters.

To create our line list, we considered 10\AA{}-wide spectral regions around strong Mn lines.
Our list of strong Mn lines was initially produced from all Mn absorption lines within the DEIMOS spectral range ($\approx$4500-6500\AA{}) from the NIST Atomic Spectra Database\footnote{The NIST Atomic Spectra Database is available at \url{https://www.nist.gov/pml/atomic-spectra-database}}.
This line list was then vetted by determining which lines were likely to be useful for distinguishing Mn abundances.

First, $10$\AA{}-wide spectral regions centered on each Mn line were synthesized and smoothed to match DEIMOS resolution. 
To determine which Mn lines would be sensitive to a $0.5$~dex change in metallicity, we estimated the relative change in line strength for each line:
\begin{equation}
\Delta(f_{\lambda}) = \frac{f_{\lambda}([\mathrm{Mn/H}] = 0) - f_{\lambda}([\mathrm{Mn/H}] = -0.5)}{f_{\lambda}([\mathrm{Mn/H}] = 0)},
\end{equation}
where $f_{\lambda}([\mathrm{Mn/H}] = X)$ denotes the flux decrement of the synthetic spectral line at $\lambda$ assuming a manganese abundance of $[\mathrm{Mn/H}] = X$.
Lines were discarded from the list if $\Delta(f_{\lambda})$ was smaller than a threshold value of $1$\%.

We further determined which Mn lines were likely to be useful by synthesizing spectra using the known Mn abundances of the Sun and of Arcturus and directly comparing each line with the observed spectra of these stars.
Any manganese absorption lines for which the amplitudes or shapes of the synthetic spectral lines were strongly inconsistent with the observed spectra were discarded.

Finally, resonance lines (lines with excitation potential $0$~eV) were removed from the line list.
These lines have been known to yield significantly lower Mn abundances compared to those measured from higher-excitation lines \citep[e.g.,][]{Bonifacio09,Sneden16}.
Resonance lines are also the most sensitive to deviations from local thermodynamic equilibrium \citep[``non-LTE (NLTE) effects''; e.g.,][]{Bergemann08,Bergemann19}.
We discuss other potential implications of non-LTE effects in Section~\ref{sec:nlte}.

\begin{deluxetable}{cc}
\tablecolumns{2} 
\tablecaption{ Manganese spectral lines. \label{tab:mn}} 
\tablehead{ 
\colhead{Wavelength} & \colhead{Excitation Potential} \\
\colhead{(\AA)} & \colhead{(eV)}
}
\startdata
4739.1 & 2.914 \\ 
4754.0 & 2.282 \\  
4761.5 & 2.953 \\  
4762.3 & 2.889 \\  
4765.8 & 2.941 \\  
4766.4 & 2.920 \\  
4783.4 & 2.300 \\  
4823.5 & 2.320 \\  
5399.5 & 3.850 \\  
5407.3 & 2.143 \\  
5420.3 & 2.143 \\  
5516.8 & 2.178 \\  
5537.7 & 2.187 \\  
6013.5 & 3.072 \\  
6016.6 & 3.075 \\  
6021.8 & 3.075 \\  
6384.7 & 3.772 \\  
6491.7 & 3.763
\enddata
\end{deluxetable}

In total we consider $18$ Mn lines, described in Table~\ref{tab:mn}.
We note that hyperfine structure (HFS) can increase the line strength at fixed abundance, producing Mn abundance corrections of up to $\sim 1.5$~dex \citep{North12}.
To account for this, we used Mn HFS lines from the database maintained by R. L. Kurucz\footnote{The Kurucz line list database is available at \url{http://kurucz.harvard.edu/linelists.html}}.

Atomic and molecular lines from other species within the 10\AA{}-wide spectral regions were taken from manually-vetted solar absorption line lists from \citet{Escala19}, with oscillator strengths tuned to match high-resolution spectra of the Sun, Arcturus, and metal-poor globular cluster stars. 
The full line list used in this work is enumerated in Table~\ref{tab:linelist}.

The other required input to the pipeline is a list of stellar parameters.
Values for these stellar parameters (effective temperature $T_{\mathrm{eff}}$, surface gravity $\log(g)$, iron-to-hydrogen ratio [Fe/H], $\alpha$-to iron ratio [$\alpha$/Fe], and microturbulent velocity $\xi$) are adopted from \citet{Kirby10} for dSph galaxies, and from \citet{Kirby16} for globular clusters.
Microturbulent velocity $\xi$ is calculated from the surface gravity using the empirical formula from \citet{Kirby09}.

\begin{deluxetable}{cccc}
\tablecolumns{4} 
\tablecaption{ Full line list. \label{tab:linelist}} 
\tablehead{ 
\colhead{Wavelength} & \colhead{Species$^{a}$} & \colhead{Excitation Potential} &  \colhead{$\log gf$} \\
\colhead{(\AA)} & & \colhead{(eV)} &
}
\startdata
  4729.019 & 26.0 & 4.073 & -1.614 \\
  4729.040 & 58.1 & 3.708 & -2.780 \\
  4729.042 & 23.0 & 2.264 & -4.909 \\
  4729.046 & 25.1 & 6.139 & -2.998 \\
  4729.049 & 68.0 & 1.069 & -0.037 \\
  4729.128 & 90.0 & 0.966 & -1.221 \\
  4729.136 & 42.0 & 2.597 & -0.785 \\
  4729.168 & 26.0 & 4.473 & -2.658 \\
  4729.186 & 20.0 & 5.049 & -4.150 \\
  4729.200 & 21.0 & 1.428 & -0.530
\enddata
\tablenotetext{a}{Atomic species are denoted using the MOOG \citep{moog} format of $Z.i$, where $Z$ is the atomic number of the element and $i$ is its ionization state.}
\tablecomments{Only a portion of Table~\ref{tab:linelist} is shown here; it is published in its entirety in the machine-readable format online.}
\end{deluxetable}

\subsubsection{Continuum normalization}

Using the input line list and stellar parameters, the automated pipeline can fit synthetic spectra to an observed spectrum.
First, the observed spectrum must be corrected for the slowly-varying global continuum.
To do this, the pipeline synthesizes a spectrum with the same stellar parameters as the observed spectrum, but with a solar Mn abundance.
The synthetic spectrum is linearly interpolated from pre-generated spectral grids as in \citet{Kirby16}.

The synthetic spectrum is then interpolated and smoothed using a Gaussian kernel to match the wavelength array and resolution of the observed spectrum.
The observed spectrum is divided by the smoothed synthetic spectrum, masking out $\pm 1$\AA{} regions around Mn lines and other regions with significant continuum fluctuations (e.g., $\pm 5$~pixel regions near the CCD chip gap, $\pm 5$\AA{} regions around the H$\alpha$, H$\beta$, and H$\gamma$ Balmer lines, $\pm 8$\AA{} regions around the strong Na D doublet at $\lambda\lambda$5890, 5896\AA{}, and any pixels with negative inverse variances).
A cubic spline is fit to the unmasked portions of this quotient with breakpoints every 150 pixels (66\AA{}).
The original observed spectrum is divided by the spline, which represents the global continuum, to obtain the continuum-normalized spectrum.

\subsubsection{Spectral synthesis and fitting}
\label{sec:fitting}

Synthetic spectra can now be produced and fit to the continuum-normalized observed spectrum.
Based on the input stellar parameters, stellar atmosphere models are linearly interpolated from the ATLAS9 grid of one-dimensional plane-parallel stellar atmosphere models \citep{Kurucz93}.
Using these stellar atmosphere models and the line lists described in Section~\ref{sec:inputs}, synthetic spectra with varying Mn abundances are produced using the spectral synthesis code MOOG \citep{moog}.
To decrease computation time, only spectral regions $\pm10$\AA{} around the Mn lines are synthesized. 

As in the continuum normalization process, these synthetic regions are interpolated and smoothed to match the observed spectrum.
The pipeline then fits the synthetic regions to the observed spectrum.
To determine the best-fitting Mn abundance, a Levenberg-Marquardt least-squares fitting algorithm is used to minimize the $\chi^{2}$ statistic, with Mn abundance as the free parameter.
This is implemented using Python's \texttt{scipy.optimize.curve\_fit} function \citep{scipy}.

\begin{figure*}[t!]
    \centering
    \epsscale{1.15}
    \plotone{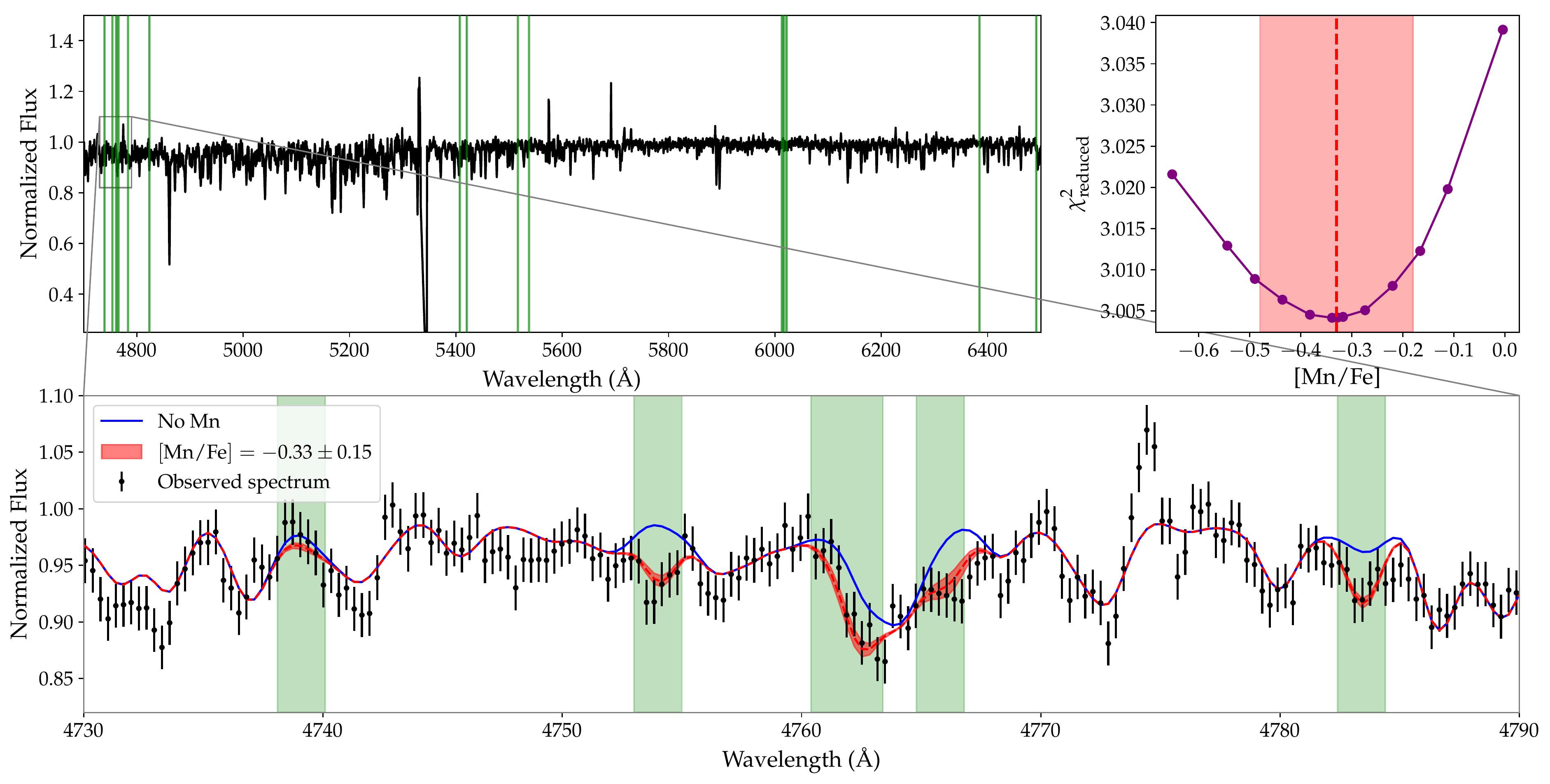}
    \caption{\textbf{(Top left)} Full continuum-normalized observed spectrum for an example star in Sculptor dSph with a high ($>84$th percentile in our sample) signal/noise ratio $S/N=67$. Green shaded regions indicate manganese lines. \textbf{(Top right)} Reduced $\chi^{2}$ as a function of [Mn/Fe]. Shaded region indicates $\pm1\sigma$ confidence interval. \textbf{(Bottom)} Zoomed-in portion of the observed spectrum (black points), again with green shaded vertical bars indicating manganese lines. The median and $\pm\sigma_{\mathrm{stat}}$ best fit is indicated by the red shaded region, while the blue line indicates a fit with negligible [Mn/Fe] ($\mathrm{[Mn/H]}= -10$) for comparison.}
    \label{fig:example}
\end{figure*}

Examples of the best-fit (continuum-normalized) spectra and reduced $\chi^{2}$ contour are shown for one star in Figure~\ref{fig:example}.
The $\chi^{2}$ contours of each star were manually inspected, and any stars whose $\chi^{2}$ contours lacked a clear minimum were removed from analysis.
Stars with a fitting error larger than $0.3$~dex (a factor of $\sim 2$) were also removed.

\subsection{Uncertainty Analysis}
\label{sec:uncertainty}
In this section, we first discuss the sources of statistical and systematic uncertainty in our measurements of [Mn/Fe]. We then validate our pipeline and assumed uncertainties by comparing our measurements of [Mn/Fe] with measurements from high-resolution spectroscopy.

\subsubsection{Statistical uncertainty}

The statistical uncertainty is dominated by the spectral noise.
This manifests in our [Mn/Fe] measurements when fitting synthetic spectra to the observed spectra, since the least-squares statistic is weighted by the uncertainties in the spectra.
The statistical uncertainty $\sigma_{\mathrm{stat}}$ is therefore given by the square root of the diagonal values of the covariance matrix, which is generated by the \texttt{scipy.optimize.curve\_fit} function.
The average statistical uncertainty in our [Mn/Fe] measurements is $\langle\sigma_{\mathrm{stat}}\rangle=0.17$~dex.

\subsubsection{Systematic uncertainty}

There are several potential sources of systematic uncertainty in our measurement pipeline. 
Uncertainties in the input stellar parameters, as well as our choice of line list, atmosphere models, and spectral synthesis code can all produce systematic errors in our [Mn/Fe] measurements.
We consider some of these sources here.

\begin{center}
\emph{Atmospheric parameter uncertainties}
\end{center}

As described in Section~\ref{sec:inputs}, our [Mn/Fe] measurements require inputs of stellar parameters $T_{\mathrm{eff}}$, $\log g$, $\xi$, [Fe/H], and [$\alpha$/Fe] in order to synthesize spectra.
We assumed fixed values of these parameters, but variations in the atmospheric parameters ($T_{\mathrm{eff}}$, $\log g$, $\xi$) may affect abundance measurements ([Fe/H], [$\alpha$/Fe], [Mn/Fe]).

We can estimate the effect of varying atmospheric parameters on our [Mn/Fe] measurement.
Since the [Mn/Fe] measurement pipeline also requires an input value of [Fe/H], we must first consider how errors in atmospheric parameters ($T_{\mathrm{eff}}$, $\log g$) may affect [Fe/H].

We note that we do not consider the effect of varying atmospheric parameters on [$\alpha$/Fe]. To some extent, measurements of [$\alpha$/H], [Mn/H], and [Fe/H] will be similarly affected by variations in the atmospheric parameters. We therefore expect that uncertainties in atmospheric parameters will contribute less significantly to errors in abundance ratios like [$\alpha$/Fe] and [Mn/Fe] than in [Fe/H].

For all stars in our sample, \citet{Kirby10} estimated the effect of varying $T_{\mathrm{eff}}$ and $\log g$ on [Fe/H].
Using these estimates, we can directly quantify systematic errors due to uncertainties in atmospheric parameters: we change $T_{\mathrm{eff}}$ by $\pm125$ and $\pm250$~K, apply the resulting changes to [Fe/H] \citep[Table 6 of][]{Kirby10}, then run our pipeline and measure the final variation in [Mn/Fe]. We repeat this procedure for $\log g$, changing $\log g$ by $\pm0.3$ and $\pm 0.6$ dex.
When varying $\log g$, we also vary microturbulent velocity $\xi$ using the calibration derived by \citet{Kirby09}:
\begin{equation}
    \xi~(\mathrm{km}~\mathrm{s}^{-1}) = 2.13 - 0.23\log g
\end{equation}

\begin{deluxetable*}{llcccc}
\tablecolumns{6} 
\tablecaption{Effect of varying atmospheric parameters on [Mn/Fe] measurements. \label{tab:stellarparams}} 
\tablehead{ 
\colhead{Object} & \colhead{ID} & \multicolumn{4}{c}{$\delta\mathrm{[Mn/Fe]}$} \\
\cline{3-6}
\colhead{} & \colhead{} & \colhead{$T_{\mathrm{eff}}\pm125$~K} &  \colhead{$T_{\mathrm{eff}}\pm250$~K} & \colhead{$\log g\pm0.3$~dex} & \colhead{$\log g\pm0.6$~dex}
}
\startdata
Scl & 1003702 & 0.02 & 0.04 & 0.02 & 0.05 \\
Scl & 1007989 & 0.01 & 0.03 & 0.01 & 0.01 \\
Scl & 1009387 & 0.00 & 0.01 & 0.02 & 0.03 \\
Scl & 1009510 & 0.01 & 0.03 & 0.02 & 0.02 \\
Scl & 1011529 & 0.02 & 0.04 & 0.04 & 0.05 \\
Scl & 1014514 & 0.02 & 0.03 & 0.02 & 0.03 \\
Scl & 1004020 & 0.02 & 0.04 & 0.02 & 0.02 \\
Scl & 1004084 & 0.03 & 0.05 & 0.01 & 0.01 \\
Scl & 1004448 & 0.01 & 0.04 & 0.04 & 0.07 \\
Scl & 1004645 & 0.03 & 0.04 & 0.01 & 0.02 \\
\enddata
\tablecomments{Only a portion of Table~\ref{tab:stellarparams} is shown here; it is published in its entirety in the machine-readable format online.}
\end{deluxetable*}

\begin{figure*}[t!]
    \centering
    \epsscale{1}
    \plottwo{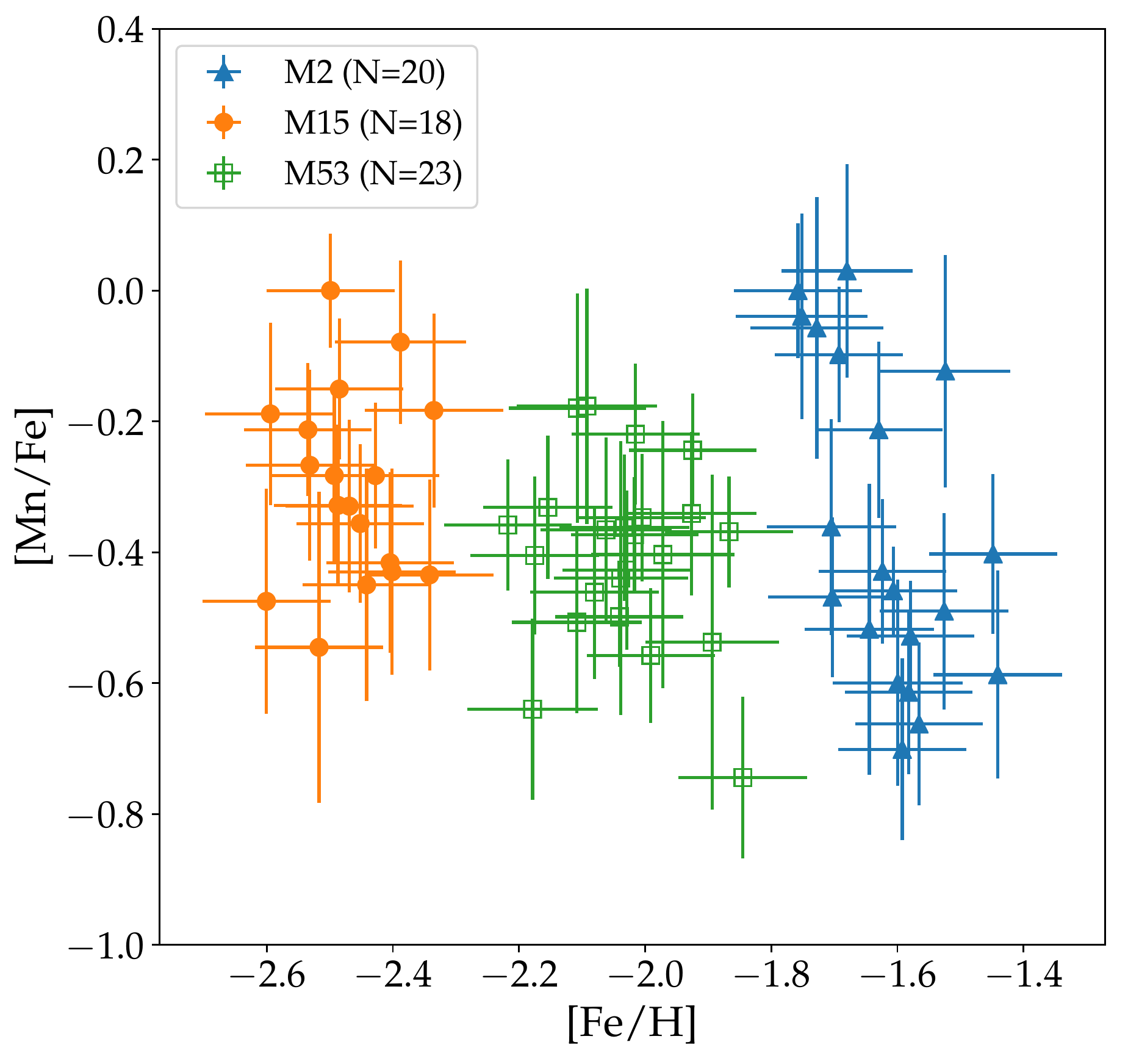}{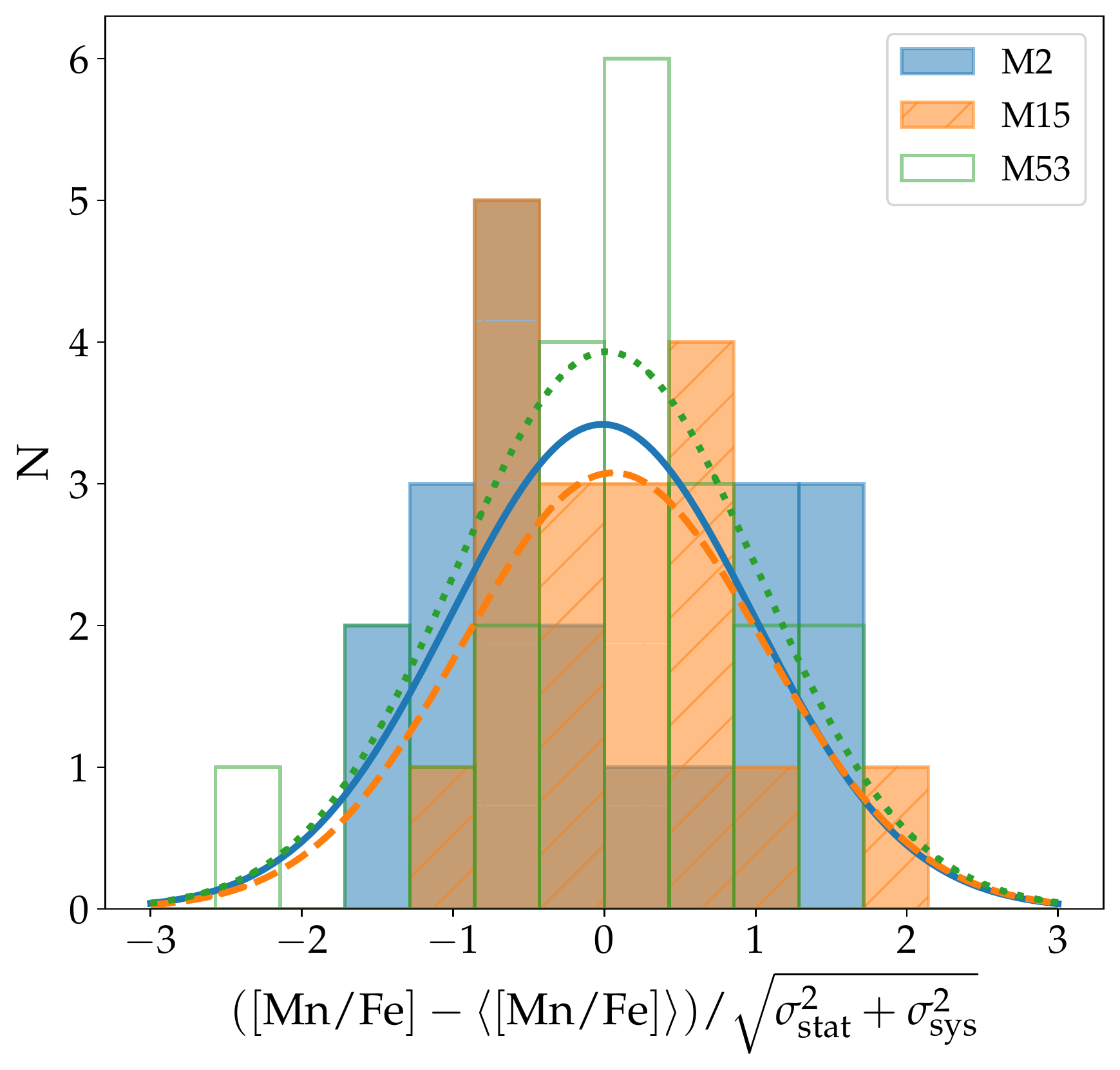}
    \caption{\textbf{(Left)} Globular cluster [Mn/Fe] abundances measured from medium-resolution spectra as a function of total metallicity [Fe/H]. \textbf{(Right)} Distribution of deviation from the mean [Mn/Fe], in units of ``total error'' (including both statistical and systematic error). Lines indicate best-fit normal distributions $\mathcal{N}(0,1)$.}
    \label{fig:gcabunds}
\end{figure*}

We report the response of [Mn/Fe] to changes in atmospheric parameters for a representative subsample of stars in Sculptor dSph, shown in Table~\ref{tab:stellarparams}.
The values listed in this table are the average absolute values of the changes in [Mn/Fe] caused by varying $T_{\mathrm{eff}}$ or $\log g$. 

The responses of [Mn/Fe] to variations in atmospheric parameters are approximately linear within $T_{\mathrm{eff}}\pm250$~K and $\log g\pm 0.6$~dex.
On average, [Mn/Fe] changes by $\pm 0.014$~dex per $\pm 100$~K change in $T_{\mathrm{eff}}$.
Similarly, [Mn/Fe] changes by $\pm 0.008$~dex per $\pm 1$~dex change in $\log g$.
These responses are relatively small compared to the average statistical error ($\langle\sigma_{\mathrm{stat}}\rangle=0.17$~dex), suggesting that any systematic errors in our [Mn/Fe] measurements due to errors in stellar parameters are negligible.
As expected, varying $T_{\mathrm{eff}}$ and $\log~g$ affects [Mn/Fe] significantly less than [Fe/H]; \citet{Kirby10} found [Fe/H] changed by $\pm 0.092$~dex per $\pm 100$~K change in $T_{\mathrm{eff}}$ and $\pm 0.039$~dex per $\pm 1$~dex change in $\log g$.

\begin{center}
    \emph{Error floor estimation using globular clusters}
\end{center}

Uncertainty in stellar parameters is unlikely to be the only source of systematic uncertainty. 
However, quantifying all individual sources of the systematic error budget is beyond the scope of this paper.
We instead estimate the value of a \emph{total} systematic error $\sigma_{\mathrm{sys}}$ by assuming globular clusters have no intrinsic dispersion in [Mn/Fe].
This $\sigma_{\mathrm{sys}}$ subsumes the error from atmospheric parameter variation discussed above, and can be added as an ``error floor'' to the statistical uncertainties to estimate final uncertainties.

To compute $\sigma_{\mathrm{sys}}$, we assume that globular clusters have little intrinsic dispersion in [Mn/Fe].
This assumption does not hold for all stellar abundances; for example, M2 (NGC 7089) appears to host two populations of stars with distinct C, N, Ba, and Sr abundances, suggesting that M2 has a complex star formation history \citep[e.g.,][]{Lardo13}. 
Similarly, M15 (NGC 7078) also displays star-to-star variation in heavy elements \citep[e.g.,][]{Sneden97}.
However, since manganese is an iron-peak element and should be formed in the same sites as iron, we expect each globular cluster to display roughly zero intrinsic dispersion in [Mn/Fe] abundance.\footnote{Some clusters do have abundance spreads in iron: $\omega$ Centauri \citep[e.g.,][]{Johnson10}, M54 \citep[][]{Carretta10}, and Terzan 5 \citep[e.g.,][]{Massari14}. However, these unusual cases are not in our sample.}

Following the procedure of \citet{Kirby10} and \citet{Duggan18}, the assumption of no intrinsic dispersion in [Mn/Fe] suggests that our measurements of [Mn/Fe] should be distributed normally about some mean $\langle [\mathrm{Mn/Fe}]\rangle$ with standard deviation equal to the combined statistical and systematic errors:
\begin{equation}
    \mathrm{stddev}\left(\frac{[\mathrm{Mn/Fe}] - \langle [\mathrm{Mn/Fe}]\rangle}{\sqrt{\sigma_{\mathrm{stat}}^{2} + \sigma_{\mathrm{sys}}^{2}}}\right) = 1
    \label{eq:sys_intrinsic}
\end{equation}
The value of $\sigma_{\mathrm{sys}}$ can then be computed from Equation~\ref{eq:sys_intrinsic}.

This calculation yields $\sigma_{\mathrm{sys}}=0.19,0.14,0.05$~dex for M2, M15, and M53 respectively.
To visualize this, the left panel of Figure~\ref{fig:gcabunds} displays the measured [Mn/Fe] abundances for these globular clusters.
The right panel of Figure~\ref{fig:gcabunds} shows distributions of deviation from the average [Mn/Fe] (i.e., $\mathrm{[Mn/Fe]} - \langle[\mathrm{Mn/Fe}]\rangle$) in units of the total error $\sqrt{\sigma_{\mathrm{stat}}^{2} + \sigma_{\mathrm{sys}}^{2}}$ for each cluster.
The distributions for M15 and M53 are well-fit by a Gaussian with a standard deviation $\sigma=1$, as expected.
M2, on the other hand, appears to have a bimodal distribution of [Mn/Fe].
This may be a result of poor membership selection; M2 has a low radial velocity \citep[$|v_{r}|\lesssim5~\mathrm{km}~\mathrm{s}^{-1}$; e.g.,][]{Baumgardt18}, so velocity selection criteria may have falsely included foreground stars as cluster members.

Based on the intrinsic dispersions of [Mn/Fe] within globular clusters M15 and M53, we estimate an average total systematic [Mn/Fe] error of $\sigma_{\mathrm{sys}}=0.10$~dex.
This total systematic error is comparable with the statistical error from fitting ($\langle\sigma_{\mathrm{stat}}\rangle\sim0.17$~dex on average).
The systematic error and statistical error are added in quadrature to obtain the total error.
We use the total [Mn/Fe] errors for the remainder of our analysis.

We note that one of the most significant systematic assumptions in our analysis pipeline is the assumption of local thermodynamic equilibrium (LTE).
Estimating non-LTE corrections for each of the stars in our sample is beyond the scope of this work, particularly since such corrections depend on both $T_{\mathrm{eff}}$ and [Mn/Fe], and are different for each Mn line.
We instead estimate the overall effect of non-LTE corrections on our results by applying a statistical correction, which we discuss later in Section~\ref{sec:nlte}.

\subsubsection{Validation with high-resolution spectroscopy comparison}

\begin{figure*}[t!]
    \centering
    \epsscale{1.15}
    \plotone{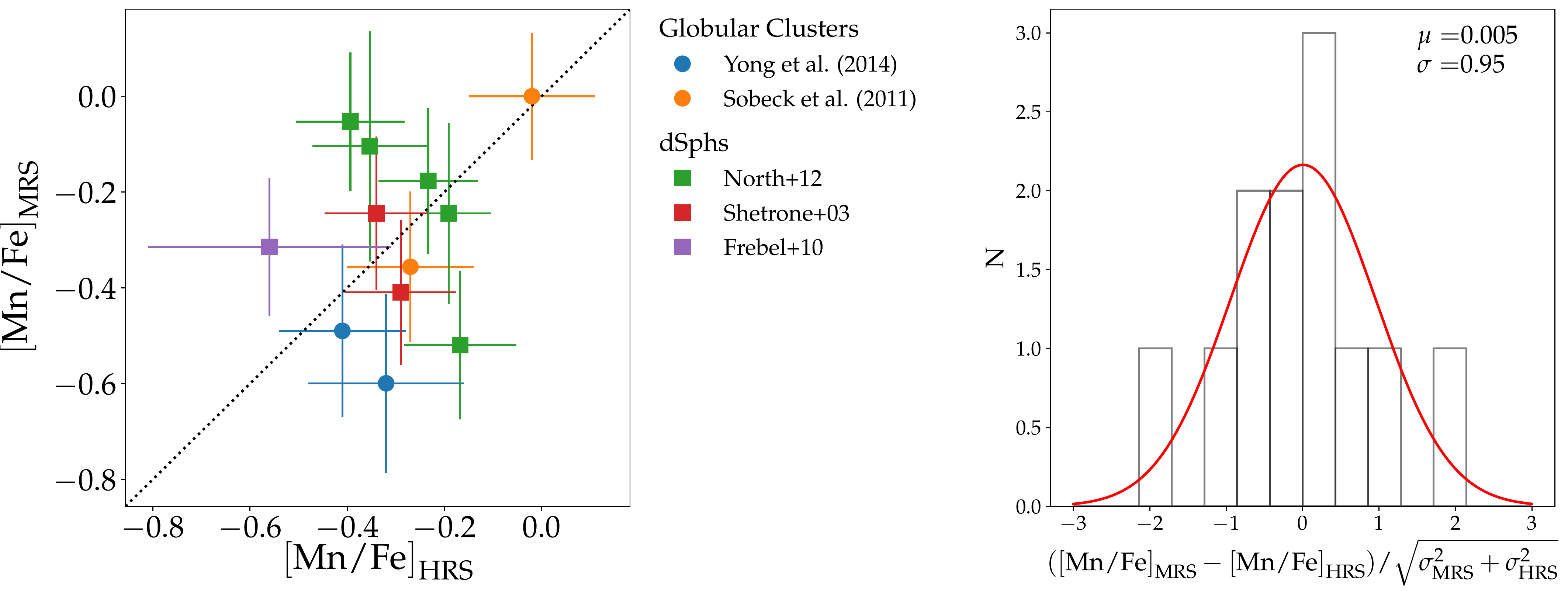}
    \caption{\textbf{(Left)} Comparison between our [Mn/Fe] measurements from medium-resolution spectra ($\mathrm{[Mn/Fe]}_{\mathrm{MRS}}$) and literature measurements from high-resolution spectra ($\mathrm{[Mn/Fe]}_{\mathrm{HRS}}$). The dotted line denotes the 1:1 line; circles (squares) denote stars from globular clusters (dSphs). \textbf{(Right)} Histogram of the differences between medium-resolution and high-resolution [Mn/Fe] measurements. The red line denotes the best-fit Gaussian distribution.}
    \label{fig:hrscomparison}
\end{figure*}

\begin{deluxetable*}{lclcc}[t!]
\tablecolumns{5} 
\tablecaption{Literature high-resolution spectroscopy catalogs. \label{tab:hrscatalogs}} 
\tablehead{ 
\colhead{Reference} & \colhead{Object} & \colhead{$N$} & \colhead{Atmospheres$^{a}$} & \colhead{Code$^{b}$} 
}
\startdata
\multicolumn{5}{c}{Globular clusters}\\[0.5em]
\tableline
\citet{Yong14} & M2 & 2 & ATLAS9 & MOOG \\
\citet{Sobeck06} & M15 & 2 & ATLAS9 & MOOG \\
\tableline
\multicolumn{5}{c}{dSphs}\\[0.5em]
\tableline
\citet{North12} & Sculptor, Fornax & 5 & MARCS & MOOG, CALRAI \\
\citet{Shetrone03} & Fornax, Leo I & 2 & MARCS & MOOG \\
\citet{Frebel10} & Ursa Major II & 1 & ATLAS9 & MOOG \\
\enddata
\tablenotetext{a}{ATLAS9: \citet{Castelli03}, \url{http://kurucz.harvard.edu/grids.html}; \\MARCS: \citet{Gustafsson75, Gustafsson03,Gustafsson08}, \url{http://marcs.astro.uu.ee}}
\tablenotetext{b}{MOOG: \citet{moog}; CALRAI: \citet{Spite67}. \citet{North12} used CALRAI for initial abundance measurements and MOOG for HFS corrections.}
\tablecomments{In all literature catalogs listed here, $T_{\mathrm{eff}}$ was measured by requiring Fe I excitation equilibrium, $\log g$ was measured by requiring Fe I and Fe II ionization balance, and $\xi$ was measured by removing abundance trends as a function of equivalent width.}
\end{deluxetable*}

\begin{deluxetable*}{lllccccccc}[t!]
\tablecolumns{10} 
\tablecaption{Comparison between DEIMOS abundances and literature high-resolution abundances. \label{tab:hrslist}} 
\tablehead{ 
\colhead{Object} & \colhead{ID} & \colhead{Reference} &  \multicolumn{5}{c}{HRS$^{b}$} & \colhead{} & \multicolumn{1}{c}{MRS} \\
\cline{4-8}
\colhead{} & \colhead{} & \colhead{} &  \colhead{$T_{\mathrm{eff}}$} & \colhead{$\log g$} & \colhead{$\xi$} & \colhead{[Fe/H]} & \colhead{[Mn/Fe]$^{a}$} & \colhead{} & \colhead{[Mn/Fe]} \\
\colhead{} & \colhead{} & \colhead{} &  \colhead{(K)} & \colhead{[cm~s$^{-2}$]} & \colhead{(km~s$^{-1}$)} & \colhead{(dex)} & \colhead{(dex)} & \colhead{} & \colhead{(dex)}
}
\startdata
M2 & An08-A1045 & \citet{Yong14} & 4275 & 0.70 & 1.78 & $-1.66$ & $-0.41\pm0.13$ & & $-0.49\pm0.18$ \\
M2 & An08-A13934 & \citet{Yong14} & 4325 & 1.30 & 1.88 & $-0.97$ & $-0.32\pm0.16$ & & $-0.60\pm0.19$ \\
M15 & 33889 & \citet{Sobeck06} & 4350 & 0.60 & 1.65 & $-2.59$ & $-0.06\pm0.13$ & & $+0.00\pm0.13$ \\
M15 & 41376 & \citet{Sobeck06} & 4225 & 0.30 & 1.85 & $-2.44$ & $-0.31\pm0.13$ & & $-0.36\pm0.16$ \\
Scl & 1008833 & \citet{North12} & \nodata & \nodata & \nodata & \nodata & $-0.27\pm0.10$ & & $-0.18\pm0.15$ \\
Scl & 1005457 & \citet{North12} & \nodata & \nodata & \nodata & \nodata & $-0.21\pm0.12$ & & $-0.52\pm0.15$\\
For & 37141 & \citet{North12} & \nodata & \nodata & \nodata & \nodata & $-0.43\pm0.11$ & & $-0.05\pm0.14$ \\
For & 54557$^{c}$ & \citet{North12} & \nodata & \nodata & \nodata & \nodata & $-0.23\pm0.09$ & & $-0.24\pm0.19$ \\
For & 67094 & \citet{North12} & \nodata & \nodata & \nodata & \nodata & $-0.39\pm0.12$ & & $-0.10\pm0.24$\\
For & 54557$^{c}$ & \citet{Shetrone03} & 4025 & 0.00 & 2.00 & $-1.21$ & $-0.40\pm0.11$ & & $-0.24\pm0.19$\\
LeoI & S60286 & \citet{Shetrone03} & 4250 & 0.80 & 2.20 &$-1.52$ & $-0.35\pm0.11$ & & $-0.41\pm0.18$ \\
UMaII & 176\_103 & \citet{Frebel10} & 4550 & 1.00 & 2.20 & $-2.34$ & $-0.56\pm0.25$ & & $-0.31\pm0.17$ \\
\enddata
\tablenotetext{a}{The errors on HRS [Mn/Fe] measurements were computed differently in each of the literature sources. However, for the most part all HRS catalogs have accounted for both statistical error (i.e., uncertainty from different Mn lines) as well as systematic error (including uncertainty from stellar parameters) in their error estimates. The only exception is the \citet{North12} HRS catalog, which does not report errors on total [Mn/Fe] abundances. For the \citet{North12} abundances, the errors listed are only the \emph{statistical} errors, estimated as the standard deviation of abundances measured from different Mn lines.}
\tablenotetext{b}{\citet{Sobeck06} obtained stellar parameters from \citet{Sneden97}. The stellar parameters used by \citet{North12} are not publicly available.}
\tablenotetext{c}{The star 54557 has two separate HRS measurements from \citet{North12} and \citet{Shetrone03}. We list them as separate entries for completeness.}
\end{deluxetable*}

We now validate our pipeline by comparing our [Mn/Fe] measurements, which are derived from medium-resolution spectra (MRS), with measurements from high-resolution spectra (HRS).
From the literature, we find $N=12$ stars in our sample that have HRS measurements; this small sample size is largely due to manganese's weak lines in the blue part of the optical wavelength range, which make it difficult to measure manganese from HRS.
In Table~\ref{tab:hrscatalogs}, we list the literature catalogs that contain HRS measurements for these 12 stars.
In Table~\ref{tab:hrslist}, we list the MRS and HRS measurements of [Mn/Fe] for these stars, as well as the stellar parameters used in the HRS measurements.

The left panel of Figure~\ref{fig:hrscomparison} compares our medium-resolution measurements ($\mathrm{[Mn/Fe]}_{\mathrm{MRS}}$) with the literature HRS measurements ($\mathrm{[Mn/Fe]}_{\mathrm{HRS}}$).
The difference between these measurements ($\mathrm{[Mn/Fe]}_{\mathrm{MRS}}-\mathrm{[Mn/Fe]}_{\mathrm{HRS}}$) is on average $-0.03$~dex.
This is significantly smaller than the median MRS and HRS errors reported for this comparison sample ($\sigma_{\mathrm{median,MRS}}\sim0.10$~dex and $\sigma_{\mathrm{median,HRS}}\sim0.16$~dex, respectively), suggesting that the MRS and HRS measurements are largely consistent.
However, there is no clear correlation between the MRS and HRS measurements, likely because our comparison sample is small and covers only a narrow range of [Mn/Fe].

Assuming that both MRS and HRS measurements have accurately estimated the total (including statistical and systematic) errors, the differences between MRS and HRS measurements ($[\mathrm{Mn/Fe}]_{\mathrm{MRS}} - [\mathrm{Mn/Fe}]_{\mathrm{HRS}}$) should be distributed normally about mean zero with standard deviation equal to the combined MRS and HRS errors ($\sqrt{\sigma_{\mathrm{HRS}}^{2} + \sigma_{\mathrm{MRS}}^{2}}$).

To check this, we plot a histogram of the differences between MRS and HRS measurements in the right panel of Figure~\ref{fig:hrscomparison}.
The best-fit Gaussian distribution to this histogram (red line) has a mean of $0.005$~dex and a standard deviation $0.95$~dex, similar to the expected normal distribution $\mathcal{N}(0,1)$.
This suggests that the total errors in our [Mn/Fe] measurements are consistent with HRS errors.

We note that many of the HRS measurements use resonance lines, which are particularly sensitive to NLTE effects \citep{Bergemann19}; as discussed in Section~\ref{sec:inputs}, we remove resonance Mn lines from our line list for that reason.
This may also contribute to systematic offsets between our MRS measurements and HRS literature measurements.
Furthermore, the HRS measurements are not a flawless comparison set; the HRS catalogs use heterogeneous measurement techniques, which may introduce additional systematic offsets among catalogs.

\subsection{Manganese Abundance Catalog}
\begin{deluxetable*}{lchhcccccc}
\tablecolumns{10} 
\tablecaption{ Manganese abundance catalog of GC and dSph stars. \label{tab:abunds}} 
\tablehead{ 
\colhead{Object} & \colhead{ID} & \nocolhead{RA} & \nocolhead{Dec} & \colhead{$T_{\mathrm{eff}}$} & \colhead{$\log g$} & \colhead{$\xi$} & \colhead{$[\alpha/\mathrm{Fe}]$} & \colhead{[Fe/H]} & \colhead{[Mn/Fe]$^{a}$} \\
\colhead{} & \colhead{} & \nocolhead{(J2000)} & \nocolhead{(J2000)} & \colhead{(K)} & \colhead{[$\mathrm{cm}~\mathrm{s}^{-2}$]} & \colhead{($\mathrm{km}~\mathrm{s}^{-1}$)} & \colhead{(dex)} & \colhead{(dex)} & \colhead{(dex)}
}
\startdata
\multicolumn{10}{c}{Globular clusters}\\[0.5em]
\tableline
M15 & 15681 & 21h29m42.91s & +12d10m57.30 & $5275\pm35$ & $+3.02\pm0.10$ & $1.43\pm0.10$ & $+0.18\pm0.10$ & $-2.39\pm0.10$ & $-0.08\pm0.16$ \\
M15 & 31227 & 21h29m56.32s & +12d09m54.80 & $4470\pm19$ & $+1.06\pm0.10$ & $1.89\pm0.06$ & $+0.23\pm0.09$ & $-2.49\pm0.10$ & $-0.33\pm0.16$ \\
M15 & 33889 & 21h29m57.17s & +12d09m42.60 & $4820\pm25$ & $+1.72\pm0.10$ & $1.73\pm0.07$ & $+0.44\pm0.09$ & $-2.50\pm0.10$ & $+0.00\pm0.13$ \\
M15 & 36569 & 21h29m57.94s & +12d10m17.00 & $4409\pm20$ & $+0.86\pm0.10$ & $1.94\pm0.06$ & $+0.22\pm0.09$ & $-2.52\pm0.10$ & $-0.55\pm0.26$ \\
M15 & 37854 & 21h29m58.30s & +12d09m54.10 & $4963\pm48$ & $+2.09\pm0.10$ & $1.65\pm0.08$ & $+0.50\pm0.10$ & $-2.59\pm0.10$ & $-0.19\pm0.17$ \\
\tableline
\multicolumn{10}{c}{dSphs}\\[0.5em]
\tableline
Scl & 1003702 & 00h59m28.72s & -33d38m56.76 & $4660\pm54$ & $+1.58\pm0.10$ & $1.77\pm0.07$ & $+0.31\pm0.14$ & $-1.95\pm0.11$ & $-0.52\pm0.22$ \\
Scl & 1007989 & 00h59m50.42s & -33d38m04.04 & $4849\pm92$ & $+2.12\pm0.10$ & $1.64\pm0.08$ & $+0.22\pm0.31$ & $-1.42\pm0.13$ & $-0.23\pm0.22$ \\
Scl & 1009387 & 00h59m56.78s & -33d39m28.45 & $4597\pm101$ & $+1.53\pm0.10$ & $1.65\pm0.08$ & $+0.01\pm0.25$ & $-1.50\pm0.21$ & $-0.43\pm0.21$ \\
Scl & 1009510 & 00h59m57.28s & -33d40m31.96 & $4677\pm57$ & $+1.76\pm0.10$ & $1.81\pm0.07$ & $+0.20\pm0.13$ & $-1.80\pm0.11$ & $-0.31\pm0.21$ \\
Scl & 1011529 & 01h00m06.35s & -33d44m57.61 & $4510\pm54$ & $+1.29\pm0.10$ & $1.72\pm0.07$ & $-0.02\pm0.15$ & $-1.48\pm0.11$ & $-0.41\pm0.15$ \\
\enddata
\tablenotetext{a}{The errors reported here are total errors (statistical and systematic errors added in quadrature). The statistical (fitting) errors can be obtained by removing the contribution from the systematic error, which we estimate (cf. Section 3.2.3) to be $\sigma_{\mathrm{sys}}=0.10~\mathrm{dex}$.}
\tablecomments{Only a portion of Table~\ref{tab:abunds} is shown here; it is published in its entirety (including coordinates) in the machine-readable format online.}
\end{deluxetable*}
Finally, we present all manganese abundances measured from medium-resolution spectra in Table~\ref{tab:abunds}.
We list here the stellar parameters $T_{\mathrm{eff}}$, $\log(g)$, [Fe/H], $[\alpha/\mathrm{Fe}]$, and $\xi$ \citep[from][]{Kirby10} used as inputs in the pipeline to measure [Mn/Fe], as well as the total error in [Mn/Fe] ($\sigma = \sqrt{\sigma_{\mathrm{sys}}^{2}+\sigma_{\mathrm{stat}}^{2}}$).

The full catalog contains manganese abundance measurements of 61 stars from 3 globular clusters and 161 stars from 6 dSph galaxies.
This is one of the largest self-consistent samples of dwarf galaxy manganese abundances measured to date. 
As previously noted, high-resolution measurements are often heterogenous in their assumptions (e.g., Table~\ref{tab:hrslist}). 
The internal consistency of this catalog makes it particularly useful for galactic archaeology studies that require statistical samples of abundances.
In the next sections, we use our sample of [Mn/Fe] abundances in dSphs for such a study.

\section{Manganese Yields in Sculptor} 
\label{sec:typeia}

\begin{figure*}[t!]
    \centering
    \epsscale{1.15}
    \plottwo{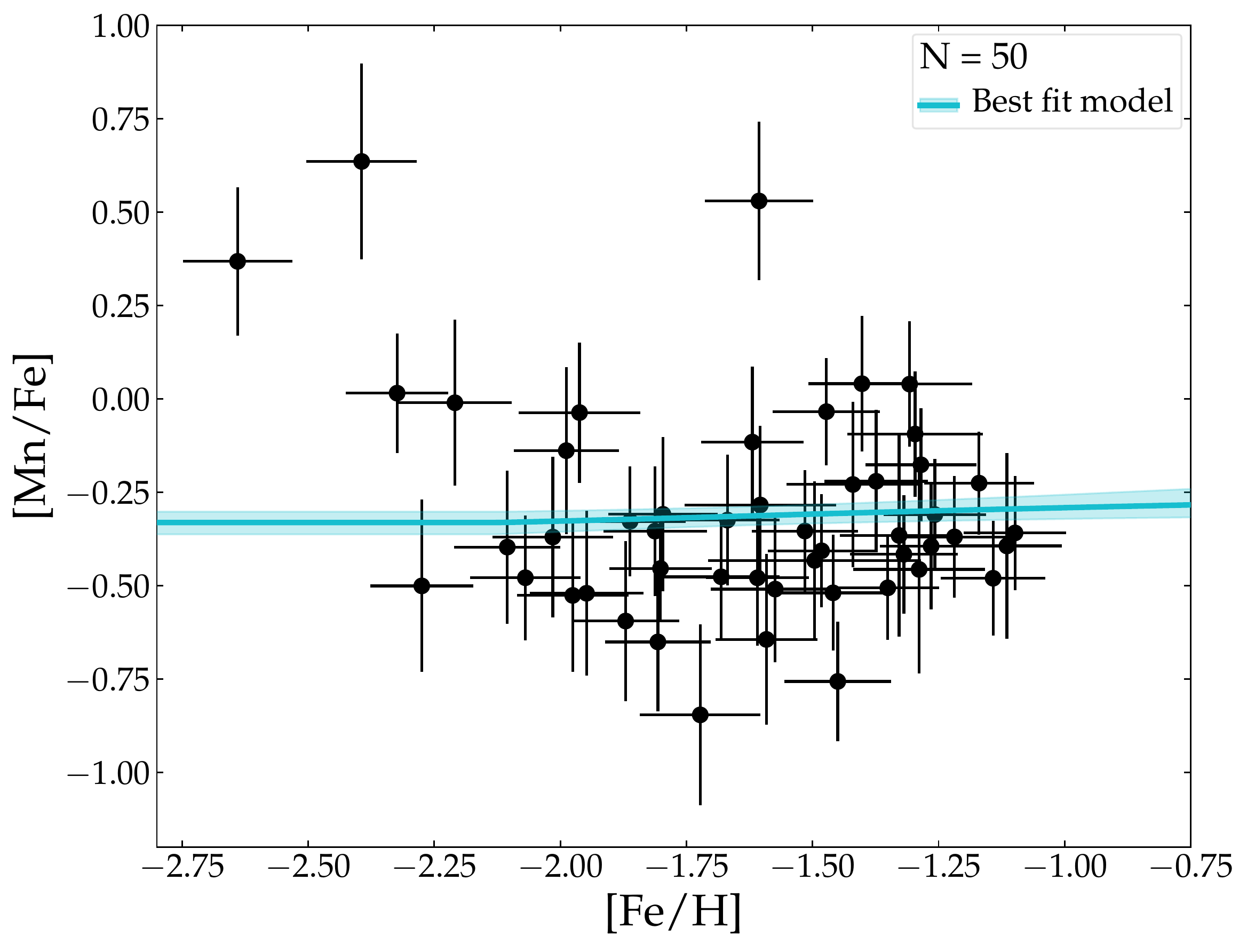}{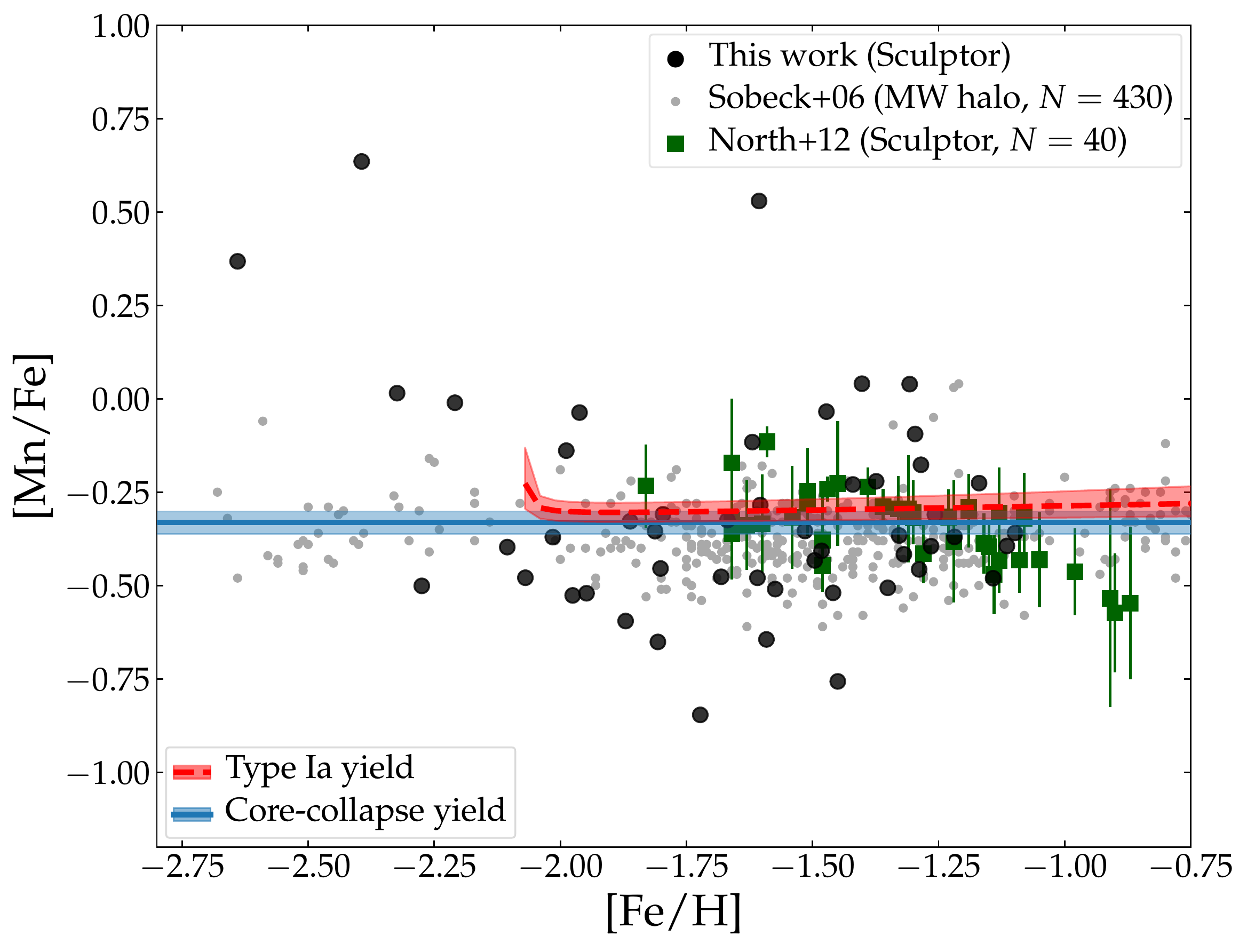}
    \caption{\textbf{(Left)} Measured [Mn/Fe] as a function of [Fe/H] for Sculptor dSph (black points). The cyan solid line marks the median best-fit model, and the cyan shaded region denotes the 68\% confidence interval about the median. \textbf{(Right)} Same, but errorbars have been removed from black points for illustration purposes. The red dashed line and shaded region marks the Type Ia [Mn/Fe] yield, and the blue solid line and shaded region marks the core-collapse supernova [Mn/Fe] yield computed from the model. Green squares denote measurements for Sculptor dSph from \citet{North12} (note that error bars only denote statistical rather than total errors); small gray points denote measurements of Milky Way halo globular cluster and field stars from \citet{Sobeck06}.}
    \label{fig:sclfit}
\end{figure*}

\subsection{Inferring [Mn/Fe] Yields from a Simple Chemical Evolution Model}
\label{sec:chemev}
With our measured manganese abundances, we can now estimate how much of this manganese is produced by Type Ia supernovae.
Following the procedure of \citet{Kirby19}, we determined Type Ia SN yields of manganese by assuming a simple chemical evolution model.
We refer readers to \citet{Kirby19} for a more detailed discussion of this model, but summarize this procedure briefly here.

This simple model assumes that core-collapse supernovae (CCSNe) are the only nucleosynthetic sources at early times, and that CCSN yields are independent of total metallicity ([Fe/H]).
The stars formed at such early times will have low [Fe/H]; furthermore, these stars will all have the same chemical abundances determined by the CCSN yields.
Put another way, for any element X, [X/Fe] will be constant as a function of [Fe/H] for low-[Fe/H] stars.

After some delay-time, Type Ia supernovae will begin to explode and produce different yields of element X.
Therefore, for stars with metallicities above some threshold $\mathrm{[Fe/H]}_{\mathrm{Ia}}$, [X/Fe] will begin to deviate from the original CCSN-only value ($\mathrm{[X/Fe]}_{\mathrm{CC}}$).
We can model this behavior with the following parameterization:
\begin{equation}
    \mathrm{[X/Fe]} = 
    \begin{cases}
    \mathrm{[X/Fe]}_{\mathrm{CC}} & \mathrm{[Fe/H]}\leq\mathrm{[Fe/H]}_{\mathrm{Ia}}\\
    \mathrm{[Fe/H]}\tan{\theta}+\frac{b_{\perp}}{\cos{\theta}} & \mathrm{[Fe/H]}>\mathrm{[Fe/H]}_{\mathrm{Ia}}
    \end{cases}
\end{equation}
where continuity is enforced at $\mathrm{[Fe/H]}=\mathrm{[Fe/H]}_{\mathrm{Ia}}$.
As described in \citet{Kirby19}, the sloped line in the $\mathrm{[Fe/H]}>\mathrm{[Fe/H]}_{\mathrm{Ia}}$ regime is parameterized by an angle ($\theta$) and perpendicular offset ($b_{\perp}$) rather than by a slope and intercept, in order to avoid biasing the linear fit toward shallower slopes \citep{Hogg10}.

Using this model, the free parameters $\mathrm{[Fe/H]}_{\mathrm{Ia}}$, $b_{\perp}$, and $\theta$ can be determined by maximizing the likelihood function $L$ \citep[Eqs. 3-6 in][]{Kirby19}.
To do the fitting, we used the \texttt{emcee} Python module \citep{Foreman-Mackey13} to minimize $-\ln L$ by implementing a Markov chain Monte Carlo (MCMC) ensemble sampler.
We ran 100 ensemble members or ``walkers,'' each with $10^{5}$ links sampled using a Metropolis-Hastings algorithm.
We discarded the first $10^{4}$ ``burn-in'' links. 

We assumed uniform priors\footnote{Specifically, we assumed $b_{\perp}\sim\mathcal{U}\{-10,10\}$ and $\theta\sim\mathcal{U}\{-\frac{\pi}{2},\frac{\pi}{2}\}$.} on $b_{\perp}$ and $\theta$, but we used the values of $\mathrm{[Fe/H]}_{\mathrm{Ia}}$ previously measured by \citet{Kirby19}\footnote{Note that \citet{Kirby19} also imposed an additional prior on $\mathrm{Mg/Fe}_{\mathrm{CC}}$, since magnesium is almost entirely produced in core-collapse supernovae.}.
As in \citet{Kirby19}, we imposed an additional prior to avoid negative values of the linear ratio $\mathrm{(Mn/Fe)}_{\mathrm{Ia}}$, which are unphysical: if any step in the MCMC chain yields $\mathrm{(Mn/Fe)}_{\mathrm{Ia}} < 0$, we set the likelihood equal to zero.
We further imposed a prior on $\mathrm{(Mn/Fe)}_{\mathrm{CC}}$:
\begin{equation}
    P = \frac{1}{\sqrt{2\pi\sigma_{\mathrm{Mn}}}}\exp\left(-\frac{(\mathrm{[Mn/Fe]}_{\mathrm{halo}}-\mathrm{[Mn/Fe]}_{\mathrm{CC}})^{2}}{2\sigma_{\mathrm{Mn}}^{2}}\right)
\end{equation}
Based on high-resolution measurements of metal-poor stars in the Milky Way halo compiled in the online database JINAbase \citep{Abohalima18}, we set $\mathrm{[Mn/Fe]}_{\mathrm{halo}}=-0.3$ and $\sigma_{\mathrm{Mn}}=0.1$.
We found that this additional prior on $\mathrm{(Mn/Fe)}_{\mathrm{CC}}$ does not significantly affect our results, since the enforced continuity at $\mathrm{[Fe/H]}_{\mathrm{Ia}}$ requires a low inferred value of $\mathrm{(Mn/Fe)}_{\mathrm{CC}}$.

The MCMC sampled the posterior distribution of the parameters $b_{\perp}$ and $\theta$.
The initial values of $b_{\perp}$ and $\theta$ were chosen by performing a simple linear fit to the [Mn/Fe] versus [Fe/H] trend for $\mathrm{[Fe/H]} > \mathrm{[Fe/H]}_{\mathrm{Ia}}$.
Unless otherwise noted, for all quantities we report the median (50th percentile) value and 68\% confidence intervals about the median.

For Sculptor, we find that [Mn/Fe] is near-constant as a function of [Fe/H], with $\theta=1.61_{-1.30}^{+2.45}$~degrees and $b_{\perp}=-0.26_{-0.05}^{+0.07}$~dex.
The data and corresponding best fit model are shown in the left panel of Figure~\ref{fig:sclfit}.
There are three high-[Mn/Fe] outliers, but removing them does not significantly change our main results, again due to the enforcement of continuity in our model. 

Using this best fit model, we can infer the CCSN and Type Ia yields of manganese from the parameters $b_{\perp}$ and $\theta$.
As described in \citet{Kirby19}, the core-collapse yield of [Mn/Fe] can be calculated as
\begin{equation}
    \mathrm{[Mn/Fe]}_{\mathrm{CC}} = \frac{b_{\perp}}{\cos\theta} + \mathrm{[Fe/H]}_{\mathrm{Ia}}\tan\theta.
    \label{eq:typeii}
\end{equation}
The Type Ia yield can then be determined from 
\begin{equation}
    \mathrm{\left(\frac{\mathrm{Mn}}{\mathrm{Fe}}\right)_{\mathrm{Ia}}} = \frac{R+1}{R}\mathrm{\left(\frac{\mathrm{Mn}}{\mathrm{Fe}}\right)_{\mathrm{\star}}} - \frac{1}{R}\mathrm{\left(\frac{\mathrm{Mn}}{\mathrm{Fe}}\right)_{\mathrm{CC}}}
    \label{eq:typeia}
\end{equation}
where $R\equiv\frac{\mathrm{Fe}_{\mathrm{Ia}}}{\mathrm{Fe}_{\mathrm{CC}}}$ is the amount of iron produced by Type Ia supernovae relative to iron produced by core-collapse supernovae.
Note that Equation~\ref{eq:typeia} does not use bracket notation, as it includes linear rather than logarithmic element ratios.

Using these equations, we compute the [Mn/Fe] yields for Sculptor.
These are denoted in the right panel of Figure~\ref{fig:sclfit} by the blue and red shaded regions, which represent the inferred CCSN and Type Ia SN yields, respectively.
We find $\mathrm{[Mn/Fe]}_{\mathrm{CC}} = -0.33_{-0.03}^{+0.03}$ for core-collapse supernovae, and $\mathrm{[Mn/Fe]}_{\mathrm{Ia}}=-0.30_{-0.03}^{+0.03}$ at $\mathrm{[Fe/H]}=-1.5$~dex for Type Ia supernovae.

Although we have manganese measurements for stars in the dSphs Ursa Minor, Ursa Major II, Canes Venatici I, Leo I, and Fornax, we do not include them in this section. 
In Ursa Minor, Ursa Major II, and Canes Venatici I, the samples of stars for which we were able to measure [Mn/Fe] are so small that we cannot draw meaningful conclusions. 
Leo I and Fornax are not well fit by our simple chemical evolution model.
We discuss these other dSphs later in Section~\ref{sec:othergalaxies}.

\subsection{Comparison with Prior Work}
\label{sec:priorwork}
We now compare our measurements with previous literature. 
The right panel of Figure~\ref{fig:sclfit} compares the Sculptor dSph manganese abundances from this work (black points) directly with those measured by \citet{North12} (green squares) and \citet{Sobeck06} (small gray points).

Our measurements imply that in Sculptor, [Mn/Fe] is roughly constant with respect to [Fe/H], suggesting that the overall manganese abundance does not change with time---and that Type Ia supernovae and core-collapse supernovae produce roughly the same yields of manganese with respect to iron.
This is consistent with \citet{North12}, who published the previously largest literature catalog of manganese abundances in Sculptor.
\citet{North12} obtained Mn abundances for $\sim40$ stars from high-resolution spectroscopy.
From their measurements, they found a plateau in [Mn/Fe] at metallicities $-1.75\lesssim\mathrm{[Fe/H]}\lesssim-1.4$, which largely agrees with our finding of metallicity-independent [Mn/Fe].

However, at a given [Fe/H], our measurements indicate a larger spread in [Mn/Fe] than \citet{North12} find.
This may be because of the different line lists used.
While we use the same 5407\AA, 5420\AA, and 5516\AA{} Mn lines that \citet{North12} use, we use also 15 other lines, including several in the bluer range of the optical spectrum ($4700-5000$\AA).
According to our line sensitivity analysis (Section~\ref{sec:inputs}) these blue lines are among the most sensitive to Mn abundance, so our measurements may be able to probe lower [Mn/Fe] than \citet{North12}, who discard any stars in their sample with ``unreliable'' Mn lines.

Furthermore, at higher metallicities \citet{North12} reported a decreasing trend of [Mn/Fe] with respect to [Fe/H].
This trend does not appear in any of the other galaxies measured in their work, although the authors noted that a similar trend has also been observed for giants and subgiants in the globular cluster $\omega$~Centauri \citep{Cunha10, Pancino11}.
\citet{North12} interpreted the decreasing trend as the result of metallicity-dependent Mn yields from Type Ia supernovae.
We are unable to confirm this downward trend at higher metallicities, since we do not observe stars with $\mathrm{[Fe/H]}\gtrsim-1.1$.

On the other hand, our observed [Mn/Fe]-[Fe/H] relation is remarkably consistent with manganese abundances measured from $\sim 200$ Milky Way cluster and field halo stars by \citet{Sobeck06}.
\citet{Sobeck06} found an average constant value of $\langle\mathrm{[Mn/Fe]}\rangle =-0.36$ for MW halo field stars, which agrees within typical uncertainties with our measured average $\langle\mathrm{[Mn/Fe]}\rangle =-0.30$.
We note that \citet{Feltzing07} reported [Mn/Fe] yields for main sequence and subgiant stars in the MW thick disk that are on average 0.15 dex higher than \citet{Sobeck06}'s measurements at $\mathrm{[Fe/H]}\sim -1$.
As \citet{North12} suggested, this slight discrepancy may be due to differences in the line lists used, or differences in NLTE corrections between giants and less evolved stars \citep[e.g.,][]{Bergemann19}.
At higher metallicities ($\mathrm{[Fe/H]}\gtrsim-1$), \citet{Feltzing07} found that [Mn/Fe] begins to increase to super-solar abundances.
This may suggest that the thick disk has a nucleosynthetic history that is distinct from the histories of the Galactic halo and Sculptor dSph.
We return to this point in Section~\ref{sec:othergalaxies}, where we discuss the potential role of SFH in driving [Mn/Fe].

\citet{Cescutti17} compiled measurements from $N\sim20$ stars from other dSphs: Ursa Minor, Sextans, and Carina.
They observed a ``butterfly''-shaped distribution of [Mn/Fe] as a function of [Fe/H], i.e., large spreads in [Mn/Fe] at $-3.5\lesssim\mathrm{[Fe/H]}\lesssim-2.0$ and $-1.75\lesssim\mathrm{[Fe/H]}\lesssim-1.0$, with a narrow spread at an intermediate metallicity ($\mathrm{[Fe/H]}\sim-2.0$).
\citet{Cescutti17} suggested that this distribution might be characteristic of a stochastic chemical evolution model with two channels: a sub-$M_{\mathrm{Ch}}$ channel and a near-$M_{\mathrm{Ch}}$ channel with relatively weak deflagrations (a ``Type Iax'' SN channel).
We do not directly compare their results with ours, since their chemical evolution model was tuned to match the metallicity distribution function of Ursa Minor.
However, we do note that the spread in our measurements ($\sigma\sim0.29$~dex, computed as the standard deviation of all [Mn/Fe] measurements in Sculptor) is roughly consistent with the spreads predicted by these stochastic models, perhaps suggesting that the chemical evolution of Sculptor dSph is also stochastic.

Finally, we briefly discuss nucleosynthetic yields measured from X-ray emission from Type Ia SN remnants (SNRs).
\citet{Yamaguchi15} compile literature manganese-to-iron ratios for three Milky Way SNRs. 
Kepler's SNR, Tycho's SNR, and 3C 397 are measured to have manganese yields of $\mathrm{[Mn/Fe]}=0.08\pm 0.17,0.22 \pm 0.20$, and $0.47 \pm 0.14$, respectively.
While these super-solar abundances are much higher than our best-fit model ($\mathrm{[Mn/Fe]}_{\mathrm{Ia}}\sim-0.3$), these SNRs are also young and likely had progenitors with near-solar metallicities, so they are not directly comparable with our measurements. 
Their super-solar abundances may be more consistent with other measurements of high-metallicity Galactic thick disk stars \citep{Feltzing07}.

\section{Implications for Type Ia Supernova Physics}
\label{sec:implications}
We now consider the implications of our measurements on Type Ia supernova physics.
We compare our observationally-inferred Type Ia supernova yield for Sculptor with yields predicted from theoretical models (Section~\ref{sec:theory}) before discussing the interpretation of [Mn/Fe] abundances in other dSph galaxies (Section~\ref{sec:othergalaxies}). 
Finally, we consider our assumption of LTE and its impact on our results (Section~\ref{sec:nlte}).

\subsection{Comparison with Theoretical Models}
\label{sec:theory}
\begin{figure}[t!]
    \centering
    \epsscale{1.15}
    \plotone{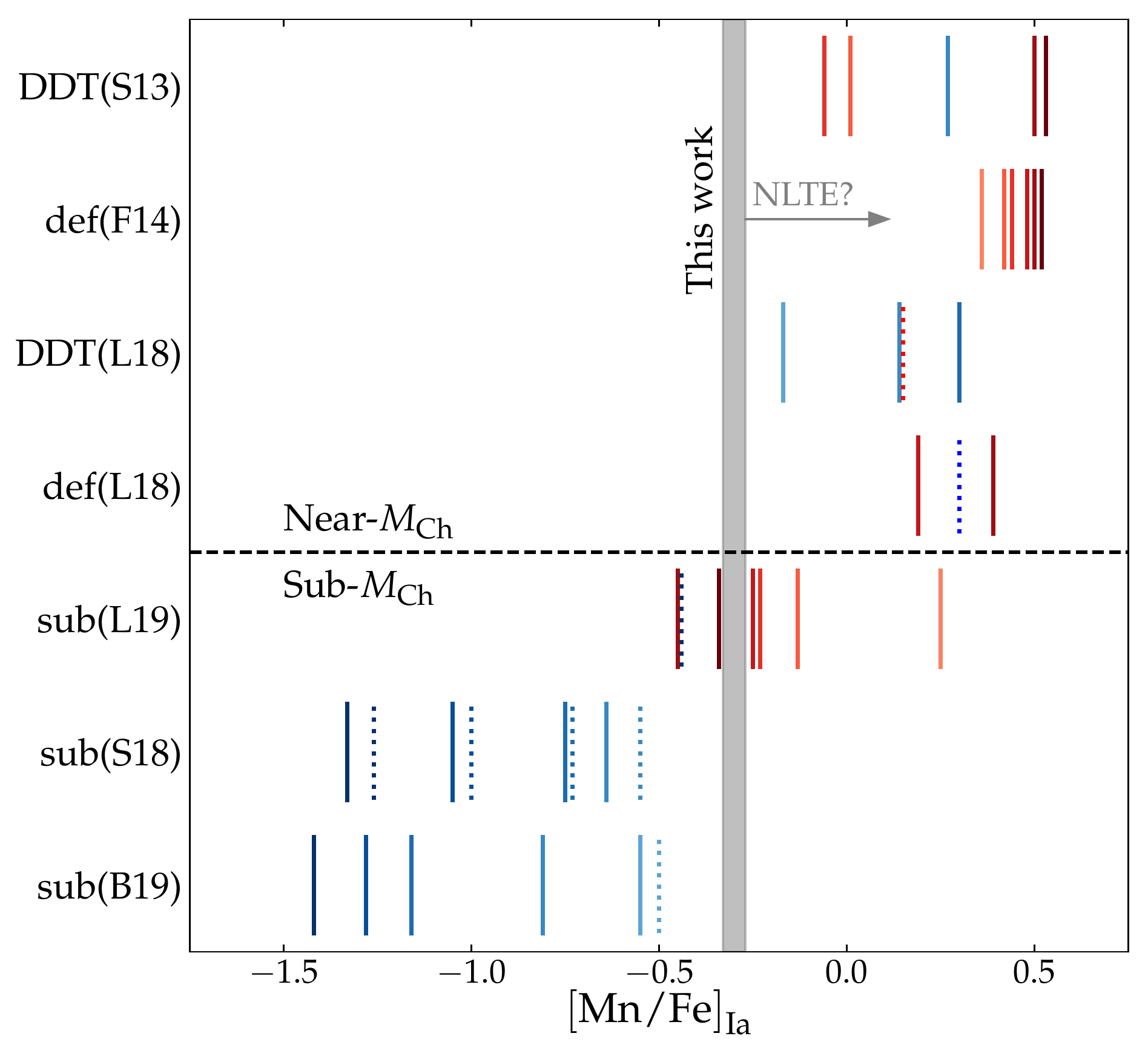}
    \caption{Type Ia supernova [Mn/Fe] yield (at $\mathrm{[Fe/H]}=-1.5$) measured in Sculptor dSph from this work (gray shaded region, marking $\pm 68\%$ confidence interval about the median), compared to theoretical yields from various models (vertical lines). Models are described in more detail in Appendix~\ref{sec:appendix}. The dashed horizontal line separates near-$M_{\mathrm{Ch}}$(above line) and sub-$M_{\mathrm{Ch}}$ models (below line). Red (blue) lines indicate theoretical yields from solar metallicity ($10^{-1.5}Z_{\odot}$) progenitors. Darker shading indicates more ignition sites (\citetalias{Seitenzahl13b} and \citetalias{Fink14}), higher initial density (\citetalias{Leung18}), or higher-mass white dwarf progenitors (\citetalias{Leung19}, \citetalias{Shen18a}, and \citetalias{Bravo19}). Dotted lines indicate special cases, denoted with asterisks in Tables~\ref{tab:typeia_yields_mch} and \ref{tab:typeia_yields_sub}.
    The gray arrow shows the maximal effect of applying NLTE corrections to our result (Section~\ref{sec:nlte}).}
    \label{fig:theorycomparison}
\end{figure}
Figure~\ref{fig:theorycomparison} compares our inferred Type Ia yield from Sculptor dSph with yields predicted from various theoretical simulations.
We discuss these models and their predicted [Mn/Fe] yields in further detail in Appendix~\ref{sec:appendix}.
We list the most relevant model details in Table~\ref{tab:theorymodels}, reproduced from Table~2 from \citet{Kirby19}.

\begin{deluxetable*}{lll}
\tablecolumns{5} 
\tablecaption{ Type Ia supernova models. \label{tab:theorymodels}} 
\tablehead{ 
\colhead{Model} & \colhead{Reference} & \colhead{Description}
}
\startdata
DDT(\citetalias{Seitenzahl13b}) & \citet{Seitenzahl13b} & ${M}_{\mathrm{Ch}}$, 3D, DDT, multiple ignition sites \\
def(\citetalias{Fink14}) & \citet{Fink14} &	${M}_{\mathrm{Ch}}$, 3D, pure deflagration, multiple ignition sites	\\
DDT(\citetalias{Leung18}) & \citet{Leung18} & ${M}_{\mathrm{Ch}}$, 2D, DDT, varying initial central density	\\
def(\citetalias{Leung18}) & \citet{Leung18} & ${M}_{\mathrm{Ch}}$, 2D, pure deflagration, varying initial central density	\\
sub(\citetalias{Leung19}) & \citet{Leung19} & sub-${M}_{\mathrm{Ch}}$, 2D, double detonation with He shell \\
sub(\citetalias{Shen18a}) & \citet{Shen18a} & sub-${M}_{\mathrm{Ch}}$, 1D, detonation of bare CO WD, two choices of C/O mass ratio \\	
sub(\citetalias{Bravo19}) & \citet{Bravo19} & sub-${M}_{\mathrm{Ch}}$, 1D, detonation of bare CO WD, two choices of ${}^{12}{\rm{C}}+{}^{16}{\rm{O}}$ reaction rate
\enddata
\tablecomments{Reproduced from Table~2 of  \citet{Kirby19}.}
\end{deluxetable*}

As discussed in Section~\ref{sec:mn}, $\mathrm{[Mn/Fe]}_{\mathrm{Ia}}$ places a strong constraint on the mass of a Type Ia progenitor.
This is shown in Figure~\ref{fig:theorycomparison}; nearly all of the near-$M_{\mathrm{Ch}}$ models (above the horizontal dashed line) produce solar or super-solar $\mathrm{[Mn/Fe]}_{\mathrm{Ia}}$, while the sub-$M_{\mathrm{Ch}}$ models can produce sub-solar $\mathrm{[Mn/Fe]}_{\mathrm{Ia}}$.
We note that when possible, we consider near-$M_{\mathrm{Ch}}$ models with $\sim1/3Z_{\odot}$ to account for core convective burning in these progenitors; we describe this ``simmering'' process further in Appendix~\ref{sec:appendix}.
We also note that the pure deflagration models def(\citetalias{Fink14}) and def(\citetalias{Leung18}) may represent near-$M_{\mathrm{Ch}}$ Type Iax supernovae.
Of the near-$M_{\mathrm{Ch}}$ models, our measured $\mathrm{[Mn/Fe]}_{\mathrm{Ia}}$ is most consistent with the low-density DDT model by \citetalias{Leung18}, which is the only near-$M_{\mathrm{Ch}}$ model to have a sub-solar [Mn/Fe] yield.\footnote{The gravitationally confined detonation of a near-$M_{\mathrm{Ch}}$ white dwarf may also have a similar sub-solar $\mathrm{[Mn/Fe]}$ yield; e.g., \citet{Seitenzahl16} find $\mathrm{[Mn/Fe]}=-0.13$ for one such model.
However, the other observables (particularly spectral features of other IMEs) predicted by this model do not match typical Type Ia SNe, and this model is therefore not expected to be a dominant channel of Type Ia SNe.} 
This model has a low central density of $1\times 10^{9}~\mathrm{g}~\mathrm{cm}^{-2}$, producing a larger detonation region which produces a very low [Mn/Fe] yield at low metallicity.
However, this central density may be unphysically low for single-degenerate Type Ia SNe \citep[e.g., Figure 4 in][]{Lesaffre06}.

Of the sub-$M_{\mathrm{Ch}}$ models, our measured Type Ia SN yield of $\mathrm{[Mn/Fe]}_{\mathrm{Ia}}=-0.30_{-0.03}^{+0.03}$ is most consistent with \citetalias{Leung19}'s solar metallicity models between $1.05-1.20~M_{\odot}$.
However, this is not a straightforward comparison, since we measure $\mathrm{[Mn/Fe]}_{\mathrm{Ia}}$ at $\mathrm{[Fe/H]}=-1.5$ rather than at solar metallicity. 
Of the remaining models, our measured $\mathrm{[Mn/Fe]}_{\mathrm{Ia}}$ is most consistent with the sub-$M_{\mathrm{Ch}}$ models of \citetalias{Shen18a} and \citetalias{Bravo19}, requiring white dwarf masses $<0.9~M_{\odot}$.
This mass constraint is lower than estimated by \citet{Kirby19}, who found that their measured yields of nickel matched Type Ia models from $\sim1.00-1.15~M_{\odot}$.
This discrepancy may simply be due to uncertainties in the theoretical yields; as Figure~\ref{fig:theorycomparison} shows, various sub-$M_{\mathrm{Ch}}$ models produce a wide range of $\mathrm{[Mn/Fe]}_{\mathrm{Ia}}$ yields due to varying physical assumptions made in the models.
Alternatively, our observationally-inferred yield may be incorrect.
The largest uncertainty in our measurement is the assumption of LTE, and we address the effect of non-LTE corrections in Section~\ref{sec:nlte}.

If we take the theoretical yields and our observationally-inferred yield at face value, then the difference between Type Ia SN models best fit by $\mathrm{[Mn/Fe]}_{\mathrm{Ia}}$ and $\mathrm{[Ni/Fe]}_{\mathrm{Ia}}$ yields must have a physical explanation.
Perhaps the most plausible explanation is that our measured yield is a combination of yields from both sub-$M_{\mathrm{Ch}}$ and near-$M_{\mathrm{Ch}}$ Type Ia or Type Iax SNe.

\begin{figure}[t!]
    \centering
    \epsscale{1.2}
    \plotone{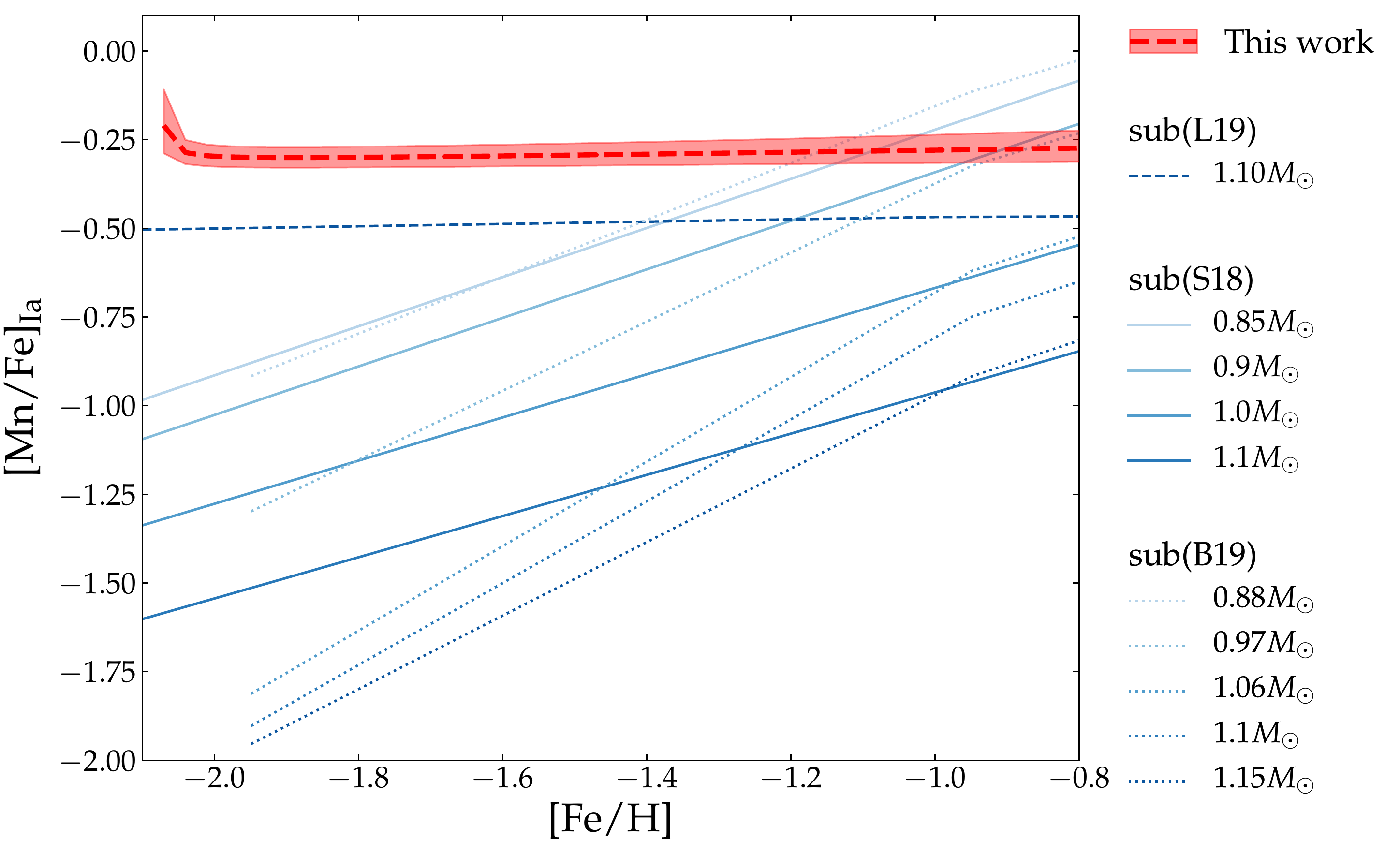}
    \caption{Type Ia SN yield of manganese $\mathrm{[Mn/Fe]}_{\mathrm{Ia}}$ as a function of [Fe/H]. The red dashed line and shaded region represent our inferred yield from Sculptor dSph, as shown in Figure~\ref{fig:sclfit}. Other lines denote yields from various sub-$M_{\mathrm{Ch}}$ theoretical models.}
    \label{fig:zdep}
\end{figure}

We further explore this hypothesis by considering the metallicity dependence of our measured $\mathrm{[Mn/Fe]}_{\mathrm{Ia}}$.
In Figure~\ref{fig:zdep}, we plot $\mathrm{[Mn/Fe]}_{\mathrm{Ia}}$ as a function of [Fe/H] and compare against theoretical predictions. 
Our observationally-inferred $\mathrm{[Mn/Fe]}_{\mathrm{Ia}}$ is near-constant as a function of metallicity across the range $-2\lesssim\mathrm{[Fe/H]}\lesssim-1$.
However, the theoretical sub-$M_{\mathrm{Ch}}$ models generally predict much larger increases in $\mathrm{[Mn/Fe]}_{\mathrm{Ia}}$ with metallicity.
This discrepancy may indicate that the combination of sub-$M_{\mathrm{Ch}}$ and near-$M_{\mathrm{Ch}}$ Type Ia SNe depends on metallicity.\footnote{This discrepancy may also be exacerbated by the dependence of $\mathrm{[Mn/Fe]}$ yields on the mass of sub-$M_{\mathrm{Ch}}$ Type Ia SNe.
More massive sub-$M_{\mathrm{Ch}}$ WDs produce lower $\mathrm{[Mn/Fe]}$ yields, and younger stellar populations should preferentially host these more massive sub-$M_{\mathrm{Ch}}$ WD explosions.
We therefore expect $\mathrm{[Mn/Fe]}_{\mathrm{Ia}}$ to be even lower at low [Fe/H].}

We can roughly estimate the fractions of sub-$M_{\mathrm{Ch}}$ and near-$M_{\mathrm{Ch}}$ Type Ia SNe required to produce our inferred $\mathrm{[Mn/Fe]}_{\mathrm{Ia}}$.
For example, at low [Fe/H], we infer a higher $\mathrm{[Mn/Fe]}_{\mathrm{Ia}}$ than sub-$M_{\mathrm{Ch}}$ SNe---particularly low-metallicity sub-$M_{\mathrm{Ch}}$ SNe---can produce.
As a conservative estimate, we consider the low-metallicity sub-$M_{\mathrm{Ch}}$ model that is least discrepant with our observed $\mathrm{[Mn/Fe]}_{\mathrm{Ia}}$: the \citetalias{Leung19} $1.1~M_{\odot}$ model.
At $\mathrm{[Fe/H]}\sim-2$, this model has a yield of $\mathrm{[Mn/Fe]}_{\mathrm{Ia}}\sim-0.50$, nearly $\sim0.3$~dex lower than our best-fit model. 
Therefore, assuming an average near-$M_{\mathrm{Ch}}$ yield from the \citetalias{Seitenzahl13b} N100 model, at least $\sim 20\%$ of SNe must be near-$M_{\mathrm{Ch}}$ SNe to reproduce our best-fit model.

If we instead compare our observationally-inferred yield with a more strongly metallicity-dependent model like those of \citetalias{Shen18a}, we can estimate the fraction of near-$M_{\mathrm{Ch}}$ Type Ia or Type Iax SNe over a range of metallicities.
Assuming $\sim1~M_{\odot}$ white dwarf progenitors as predicted by \citet{Kirby19}, we find that using \citetalias{Shen18a}'s models, $\sim 33\%$ of Type Ia SNe at $\mathrm{[Fe/H]}\sim-1$ and $\sim 36\%$ of Type Ia SNe at $\mathrm{[Fe/H]}\sim-2$ must be near-$M_{\mathrm{Ch}}$.
These estimates are somewhat higher than the fractions inferred from \citet{Kirby19}'s [Ni/Fe] measurements; using the \citetalias{Shen18a} $1~M_{\odot}$ model yields for [Ni/Fe], only $\sim22\%$ of Type Ia SNe must be near-$M_{\mathrm{Ch}}$.

We emphasize that these fractions are only rough estimates, subject to uncertainties in both the observational and theoretical yields.
However, our data suggest that some non-negligible fraction of Type Ia SNe must have near-$M_{\mathrm{Ch}}$ progenitors over the metallicity range $-2\lesssim\mathrm{[Fe/H]}\lesssim-1$.
Furthermore, the near-$M_{\mathrm{Ch}}$ fraction does not appear to change significantly across the metallicity range probed by our observations.
This may change at higher metallicities ($\mathrm{[Fe/H]}\gtrsim-1$), where 
near-$M_{\mathrm{Ch}}$ Type Ia SNe may begin to dominate, producing super-solar yields of [Mn/Fe] that are seen in, e.g., the Milky Way thick disk \citep{Feltzing07}.

As pointed out by \citet{Kirby19}, our conclusions are valid only for Type Ia SNe that occurred while Sculptor was forming stars. 
Sculptor formed the middle two thirds of its stars in 1 Gyr \citep{Weisz14}.
Our measurements are therefore sensitive to models of Type Ia SNe that have ``standard'' delay-times $<0.6$~Gyr \citep[e.g.,][]{Maoz14}. 
However, our conclusions do not account for Type Ia SNe that are delayed by more than $1$~Gyr.
Measurements of other dSphs with different star formation histories may be required to sample different varieties of Type Ia SNe, which may have longer delay-times.
We discuss this further in the next section.

\subsection{Other dSph Galaxies}
\label{sec:othergalaxies}

\begin{figure*}[t!]
    \centering
    \epsscale{1.15}
    \plottwo{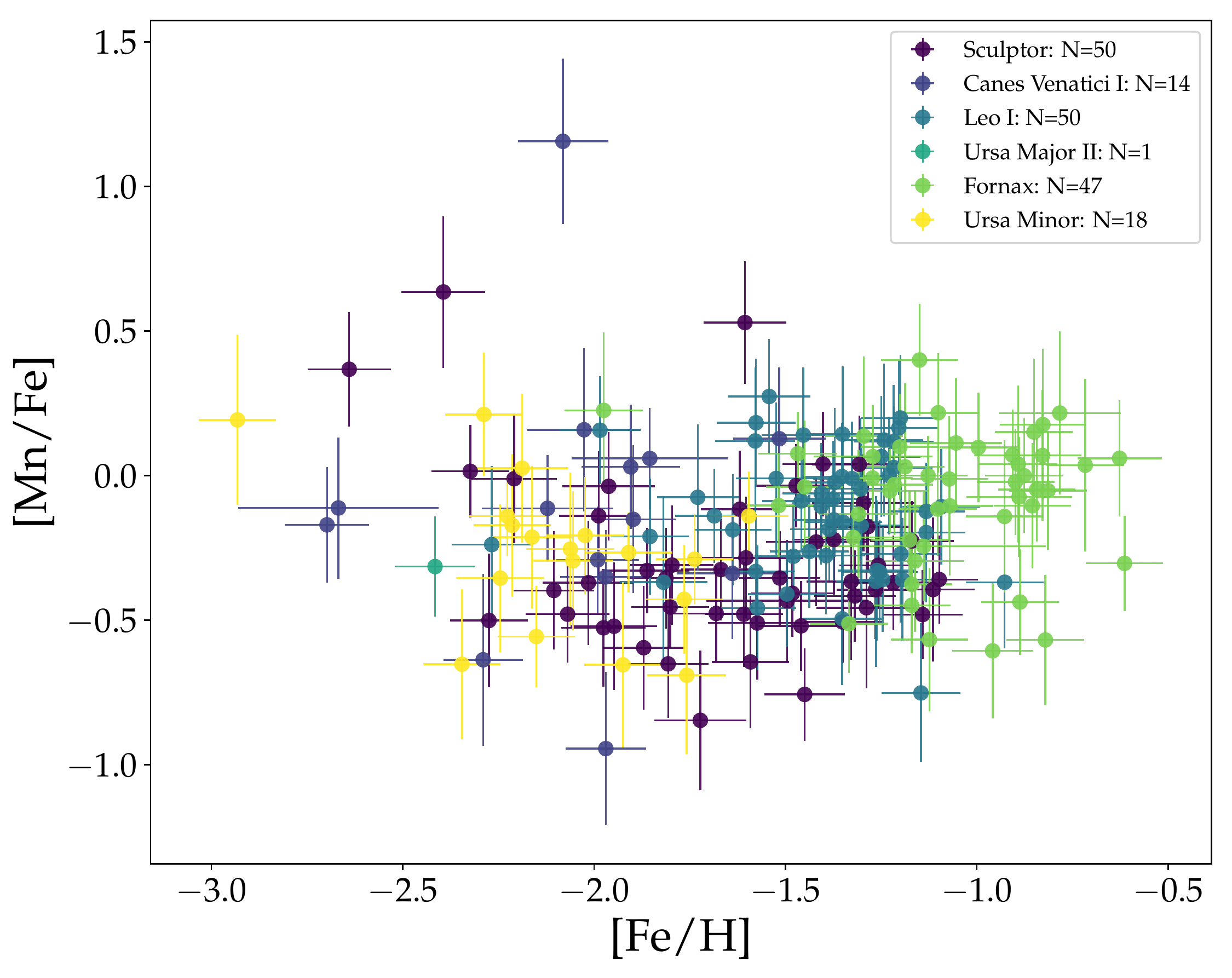}{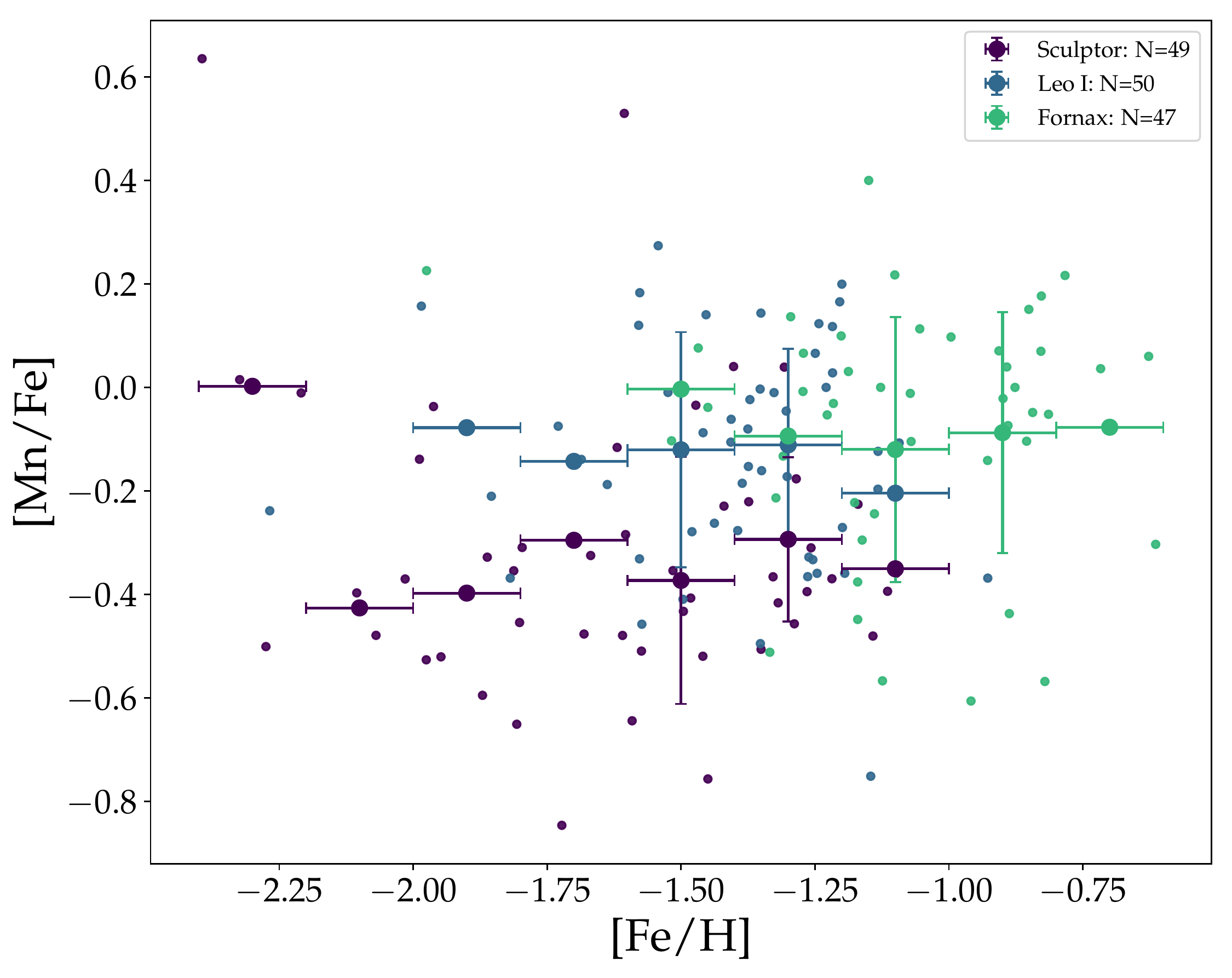}
    \caption{\textbf{(Left)} [Mn/Fe] as a function of [Fe/H] for all dSph galaxies in our sample. \textbf{(Right)} Same, but zoomed in to show only stars from Sculptor, Leo I, and Fornax dSphs. Small points denote the measured abundances (errorbars have been removed for ease of visualization), and large points with errorbars denote the weighted averages in each 0.2~dex metallicity bin (only bins with $>1$ stars are plotted, and error bars indicate combined errors of averages).}
    \label{fig:otherdsphs}
\end{figure*}

As described in Section~\ref{sec:chemev}, we are unable to fit our simple chemical evolution model to several dSphs.
Ursa Minor, Ursa Major II, and Canes Venatici have small sample sizes; Leo I and Fornax dSphs have larger sample sizes ($N=50$ and $N=45$, respectively), but are not well fit by the model.
For completeness, we illustrate the manganese abundances as a function of metallicity for all dSphs in the left panel of Figure~\ref{fig:otherdsphs}.

The right panel of Figure~\ref{fig:otherdsphs} zooms in on [Mn/Fe] as a function of [Fe/H] for the galaxies with sample sizes $N>20$: Sculptor, Leo I, and Fornax.
This illustrates that at a given [Fe/H], stars in Leo I and Fornax have higher [Mn/Fe] abundances than stars in Sculptor by $\gtrsim0.2$~dex on average.

The most obvious differences among these galaxies that might explain this discrepancy are the galaxies' star formation histories (SFHs).
Leo I and Fornax have extended star formation histories, while Sculptor's SFH is characterized by a burst of early star formation followed by a long period of low star formation rates.
This may explain the difference in [Mn/Fe] between these galaxies.
Here, we consider two potential reasons why SFH might be linked with [Mn/Fe] abundance.

First, differences in [Mn/Fe] at a given [Fe/H] may result from a combination of star formation timescales and metallicity-dependent Type Ia supernova yields.
Star formation timescales are relevant because this work uses \emph{stellar} abundances, which trace the level of chemical enrichment at the time of star formation rather than the current level of enrichment.
Thus, a star with $\mathrm{[Fe/H]}\sim-1.50$ may actually be sampling yields produced by Type Ia SNe with $\mathrm{[Fe/H]}<-1.50$ progenitors.
This ``lag'' in metallicity would be larger in Leo I and Fornax than in Sculptor, because of their extended SFHs.
If [Mn/Fe] yields from sub-$M_{\mathrm{Ch}}$ Type Ia supernovae were metallicity-dependent---more specifically, if [Mn/Fe] yields were to increase as progenitor [Fe/H] increases---then the difference in [Mn/Fe] at a given [Fe/H] between Sculptor and Fornax/Leo I might simply be a result of the difference in ``lag metallicity.''

Although a full test of this hypothesis is beyond the scope of this work, to first order we can estimate the effect of this ``lag'' by computing the average delay-time for Type Ia supernovae in each dSph.
We do this by assuming a power-law delay-time distribution \citep{Maoz12}:
\begin{equation}
    \Psi = 10^{-3}\left(\frac{t}{\mathrm{Gyr}}\right)^{-1.1}~\mathrm{SNe}~\mathrm{Gyr}^{-1}~M_{\odot}^{-1}
    \label{eq:dtd}
\end{equation}
We assume that this delay-time distribution is valid only for times later than some minimum time $t_{\mathrm{min}}\sim0.1$~Gyr.
We can then compute the average delay-time for Type Ia supernovae between $t_{\mathrm{min}}$ and some typical star formation time $t_{\mathrm{*}}$:
\begin{equation}
    t_{\mathrm{delay}} = \frac{\int_{t_{\mathrm{min}}}^{t_{*}}t\Psi\mathrm{d}t}{\int_{t_{\mathrm{min}}}^{t_{*}}\Psi\mathrm{d}t}
\end{equation}

\citet{Weisz14} find that Sculptor formed most of its stars in $\sim1$~Gyr, while Leo I and Fornax have been forming stars steadily over at least $\sim10$~Gyr.
We therefore assume an average star formation time of $t_{*}\sim0.5$~Gyr for Sculptor and $t_{*}\sim5$~Gyr for Leo I and Fornax, which yield estimates of $t_{\mathrm{delay}}\sim0.24$~Gyr and $t_{\mathrm{delay}}\sim1.12$~Gyr, respectively.
Using the age-metallicity relation for Sculptor \citep{deBoer12}, we find that this average delay-time corresponds to a metallicity lag of $\Delta\mathrm{[Fe/H]}\sim0.05$~dex; similarly, the age-metallicity relation for Fornax \citep{Letarte10} yields a metallicity lag of $\Delta\mathrm{[Fe/H]}\sim0.15$~dex.
Therefore, the difference in metallicity lags between Sculptor and Fornax is $\sim0.1$~dex.
For the \citetalias{Shen18a} $1~M_{\odot}$ Type Ia SNe model, a $\sim0.1$~dex difference in metallicity lags produces a difference in [Mn/Fe] of $\Delta\mathrm{[Mn/Fe]}\sim0.09$~dex.
This is not enough to explain the $\gtrsim0.2$~dex difference in [Mn/Fe] between Sculptor and Fornax.

Alternatively, the discrepancy in [Mn/Fe] may result from a change over time in the underlying physical mechanism behind Type Ia supernovae.
Both Leo I and Fornax have stars with significantly supersolar [Mn/Fe] abundances ($\mathrm{[Mn/Fe]}\gtrsim 0.2$~dex); as Figure~\ref{fig:theorycomparison} shows, low-metallicity sub-$M_{\mathrm{Ch}}$ Type Ia progenitors do not produce such high [Mn/Fe] yields.
This suggests that near-$M_{\mathrm{Ch}}$ white dwarf explosions may become the dominant channel for Type Ia supernovae at late times in a galaxy's star formation history.

Such a scenario---where near-$M_{\mathrm{Ch}}$ Type Ia or Type Iax supernovae explode later
than sub-$M_{\mathrm{Ch}}$ Type Ia SNe---has been proposed by, e.g., \citet{Kobayashi09}, who argue that near-$M_{\mathrm{Ch}}$ Type Ia SNe are suppressed at low metallicities due to metallicity-dependent white dwarf winds.
This scenario may also be consistent with near-$M_{\mathrm{Ch}}$ explosions requiring mass growth by hydrogen accretion, which may require a longer delay-time or higher metallicity progenitors than sub-$M_{\mathrm{Ch}}$ double degenerate mergers \citep[e.g.,][]{Ruiter11}.
As noted in Section~\ref{sec:priorwork} this also agrees with observations in the Milky Way, which show that stars in the Galactic halo have sub-solar [Mn/Fe] at $\mathrm{[Fe/H]}\lesssim-1.0$, compared to stars in the higher-metallicity thick disk, which have have super-solar [Mn/Fe] \citep{Feltzing07}.
Like Sculptor, the Milky Way halo formed most of its stars in a short early burst, while the thick disk has a more extended SFH.

\subsection{Non-LTE effects}
\label{sec:nlte}
Throughout our analysis, we have used [Mn/Fe] abundance measurements that rely on the assumption of local thermodynamic equilibrium (LTE).
In LTE, opacity is only a function of temperature and density.
However, this is only valid at high densities, when the radiation field is strongly coupled to the matter.
Previous works find that accounting for non-LTE (NLTE) effects may systematically increase Mn abundances by as much as $0.5-0.7$~dex using 1D NLTE models \citep[e.g.,][]{Bergemann08}, or up to $\sim 0.4$~dex using 3D NLTE models \citep[e.g.,][]{Bergemann19}.
We must therefore consider the effect of NLTE corrections on our results.

We estimate this by using the corrections determined by \citet{Bergemann08}, who compared Mn abundances measured using 1D LTE models and 1D NLTE models over a range of metallicities.
From Figure~9 of \citet{Bergemann19}, we find that for a typical RGB star ($T_{\mathrm{eff}}=6000$~K, $\log g=1.5$), 1D NLTE corrections ($\Delta_{\mathrm{NLTE}}=\mathrm{[Mn/Fe]}_{\mathrm{NLTE}}-\mathrm{[Mn/Fe]}_{\mathrm{LTE}}$) determined from optical lines used in this work range from $\Delta_{\mathrm{NLTE}}\lesssim0.462$~dex at $\mathrm{[Fe/H]}=-3$ to $\Delta_{\mathrm{NLTE}}\gtrsim0.173$~dex at $\mathrm{[Fe/H]}=0$.
By linearly interpolating between these bounds, we can determine a maximum ``statistical'' NLTE correction as a function of [Fe/H]\footnote{We note that although we consider primarily NLTE effects on Mn I lines, NLTE conditions can also affect Fe I lines. Predictions for NLTE Fe I corrections can be quite large \citep[up to 0.5 dex; see, e.g.,][]{Mashonkina2019, Bergemann2017, Bergemann2012}. However, these large corrections are generally applicable for metal-poor stars with $[\mathrm{Fe/H}]\lesssim -2.0$. For cool giant stars with metallicities comparable to the bulk of our sample ($[\mathrm{Fe/H}]>-2.0$), \citet{Mashonkina2019} predict NLTE corrections $\lesssim 0.1$~dex (cf. their Fig 8). This change is smaller than the average NLTE corrections predicted by Equation~\ref{eq:nlte}.}:
\begin{equation}
    \Delta_{\mathrm{NLTE}}(\mathrm{[Fe/H]}) = -0.10\mathrm{[Fe/H]} + 0.17
    \label{eq:nlte}
\end{equation}

\begin{figure}[t!]
    \centering
    \epsscale{1.2}
    \plotone{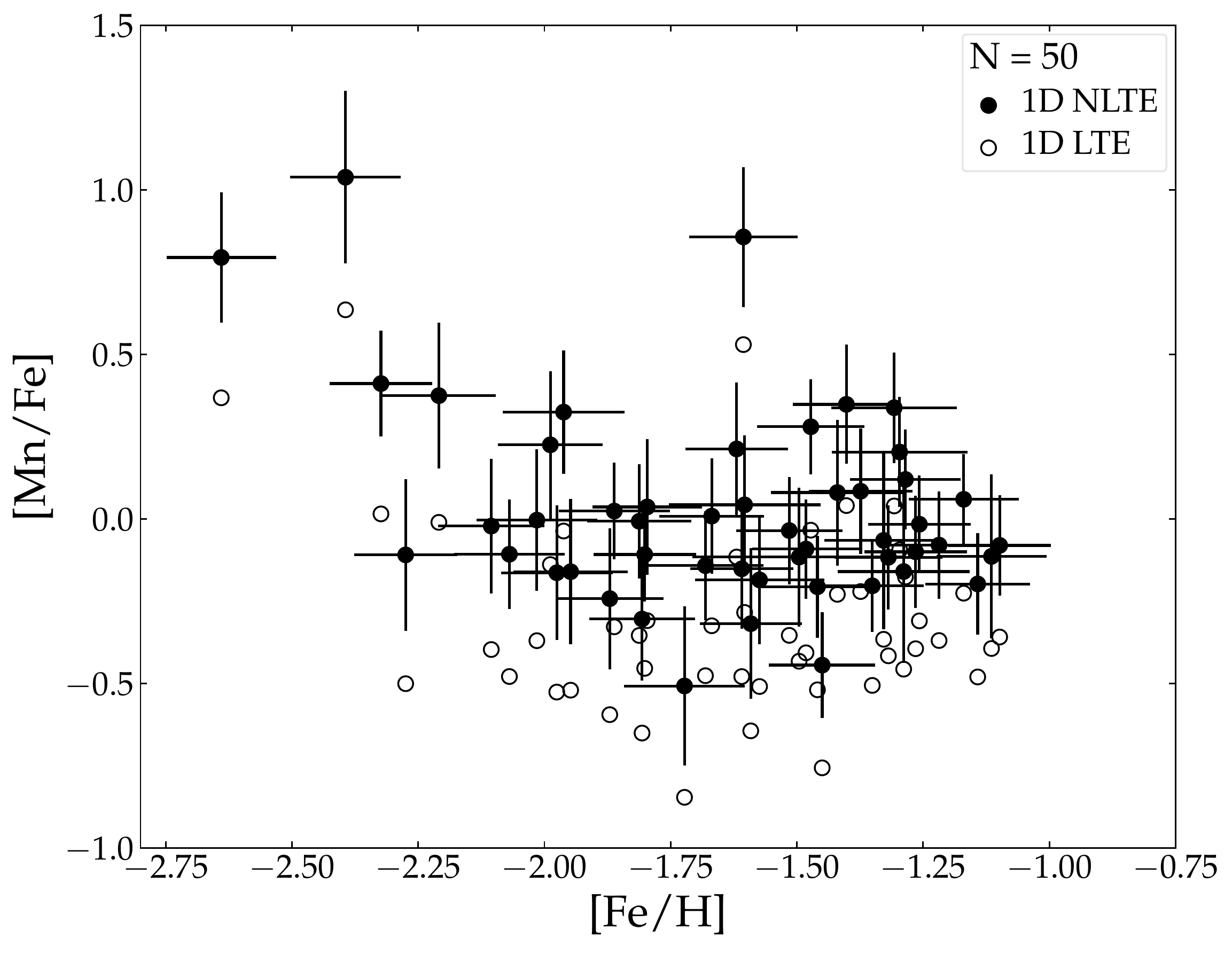}
    \caption{[Mn/Fe] as a function of [Fe/H] in Sculptor dSph. Filled points indicate the measurements with a statistical correction for 1D NLTE effects; empty points indicate the original 1D LTE measurements.}
    \label{fig:nlte}
\end{figure}

Figure~\ref{fig:nlte} shows the results of applying this maximum correction to our [Mn/Fe] measurements for stars in Sculptor dSph.
The 1D NLTE corrections have a very slight metallicity dependence, but their primary effect is to increase all of the [Mn/Fe] yields by a factor of $\sim 0.33$~dex on average.
This naturally increases the [Mn/Fe] yields inferred from Sculptor dSph: $\mathrm{[Mn/Fe]}_{\mathrm{CC,NLTE}} = 0.00_{-0.03}^{+0.03}$ for core-collapse supernovae, and $\mathrm{[Mn/Fe]}_{\mathrm{Ia,NLTE}}=+0.03_{-0.03}^{+0.03}$ at $\mathrm{[Fe/H]}=-1.5$~dex for Type Ia supernovae. 
This near-solar Type Ia yield is consistent with $M_{\mathrm{Ch}}$ theoretical models (cf. Figure~\ref{fig:theorycomparison}), a significant departure from our finding in Section~\ref{sec:theory} that the sub-$M_{\mathrm{Ch}}$ channel dominates in Sculptor dSph.
Furthermore, \citet{Bergemann19} suggest that three-dimensional effects, such as convection, may further increase [Mn/Fe] abundances in RGB stars by another $\sim0.2$~dex, producing an even higher $\mathrm{[Mn/Fe]}_{\mathrm{Ia}}$ yield.

This is an interesting difference with respect to our LTE estimates.
However, we will leave this complex analysis including detailed NLTE to a future study\footnote{This requires a complete reanalysis of stellar parameters of our targets using NLTE models.}, since \citet{Kirby18} found that applying 1D NLTE corrections instead increased the dispersion of iron-peak abundances ([Co/Fe] and [Cr/Fe]) in globular clusters.
\citet{Kirby18} suggested that this behavior is due to the method which the atmospheric parameters were determined (1D LTE modelling of spectra with a micro-turbulence relationship calibrated on LTE results).

In any case, the NLTE corrections do not appear to strongly affect the metallicity dependence of [Mn/Fe]; as in the LTE case, we observe a nearly-flat trend of [Mn/Fe] vs [Fe/H] across the metallicity range $-2.25<\mathrm{[Fe/H]}<-1.0$ in Sculptor.
Furthermore, our comparison between Sculptor and other dSph galaxies (Leo I, Fornax) depends primarily on relative differences between [Mn/Fe] abundances at a given [Fe/H].
Effective temperature might also affect the magnitude of NLTE corrections, but $T_{\mathrm{eff}}$ at a given [Fe/H] in Sculptor, Leo I, and Fornax are offset by $200-300$~K at most; the resulting difference in NLTE corrections is $\lesssim 0.05$~dex, not enough to explain the discrepancy in [Mn/Fe] at a given [Fe/H] between these galaxies.
NLTE corrections are therefore unlikely to affect our interpretation of [Mn/Fe] abundances as a function of SFH (Section~\ref{sec:othergalaxies}).

\section{Summary and Conclusions}
\label{sec:summary}

We have presented the results of medium-resolution spectra from the new 1200B grating on Keck DEIMOS.
Using a pipeline that generates synthetic stellar spectra, we have measured manganese abundances for $N=161$ stars in six classical dSph galaxies.
These manganese abundance measurements were validated using the internal dispersions of globular clusters and comparison with high-resolution spectroscopy.

By fitting a simple chemical evolution model to measurements of [Mn/Fe] as a function of [Fe/H], we have inferred the manganese yields of core-collapse and early Type Ia supernovae in Sculptor dSph:
$\mathrm{[Mn/Fe]}_{\mathrm{CC}}=-0.33^{+0.03}_{-0.03}$ and $\mathrm{[Mn/Fe]}_{\mathrm{Ia}}=-0.30^{+0.03}_{-0.03}$ (at $\mathrm{[Fe/H]}=-1.5$), respectively.
Since only sub-$M_{\mathrm{Ch}}$ Type Ia SN models are able to produce significantly sub-solar values of $\mathrm{[Mn/Fe]}_{\mathrm{Ia}}$, we conclude that the dominant explosion mechanism of Type Ia SNe that occurred before the end of star formation in Sculptor is the detonation of a sub-$M_{\mathrm{Ch}}$ WD.
However, in order to reproduce our observationally-inferred $\mathrm{[Mn/Fe]}_{\mathrm{Ia}}$, we find that a fraction ($\gtrsim20\%$) of all Type Ia SNe in our metallicity range $-2<\mathrm{[Fe/H]}<-1$ must have near-$M_{\mathrm{Ch}}$ progenitors.

This conclusion may not hold for other environments.
In particular, the Milky Way thick disk and dSphs with extended SFHs display different trends of [Mn/Fe] as a function of metallicity.
We find that at a given metallicity, dSphs with extended SFHs like Fornax and Leo I have $\gtrsim0.2$~dex higher average [Mn/Fe] abundances than Sculptor, which has an ancient SFH.
This discrepancy is large enough to imply a physical change in the nucleosynthetic source of Mn---perhaps the dominant channel of Type Ia SNe evolves over time, and near-$M_{\mathrm{Ch}}$ white dwarf detonations become the dominant channel at longer delay-times.

Finally, we consider the effect of non-LTE corrections on our results. 
Including a statistical NLTE correction increases the [Mn/Fe] yields from both core-collapse and Type Ia supernovae by $\sim0.3$~dex.
The resulting $\mathrm{[Mn/Fe]}_{\mathrm{Ia}}$ is approximately solar at $\mathrm{[Fe/H]}\sim-1.5$, more consistent with yields from near-$M_{\mathrm{Ch}}$ models.
The detailed treatment of NLTE effects, however, requires a full re-analysis of stellar parameters of our targets with NLTE synthetic spectral models. 
This will be the subject of the future work.

We also hope to test the results of this work using more data in dSphs.
Other dSphs with ancient SFHs similar to Sculptor (e.g., Draco, Canes Venatici II) could be used to confirm whether sub-$M_{\mathrm{Ch}}$ explosions dominate at early times in dwarf galaxies.
Dwarf spheroidal galaxies with diverse star formation histories, such as Carina \citep[e.g.,][]{Hernandez00}, may also be particularly intriguing environments in which to test our conclusions about the SFH dependence of Type Ia supernovae.

\acknowledgments
{The authors thank I. Escala and G. Duggan for informing parts of the data pipeline, as well as A. Piro, A. McWilliam, and the anonymous referee for helpful discussions and comments.
This material is based upon work supported by the National Science Foundation under grant No. AST-1847909. 
MAdlR acknowledges the financial support of the NSF Graduate Research Fellowship Program.
ENK gratefully acknowledges support from a Cottrell Scholar award administered by the Research Corporation for Science Advancement as well as funding from generous donors to the California Institute of Technology.
IRS was supported by the Australian Research Council through Grant No. FT160100028.

The authors wish to recognize and acknowledge the deep cultural role and reverence that the summit of Maunakea has always had within the indigenous Hawaiian community.  
We are most fortunate to have the opportunity to conduct observations from this sacred mountain.
Finally, we would like to express our deep gratitude to the staff at academic and telescope facilities, particularly those whose communities are excluded from the academic system, but whose labor maintains spaces for scientific inquiry.}

\vspace{5mm}
\facilities{Keck:II (DEIMOS)}

\software{
spec2d \citep{Cooper12,Newman13},
Matplotlib \citep{matplotlib}, 
Astropy \citep{astropy},
Scipy \citep{scipy},
MOOG \citep{moog},
ATLAS9 \citep{Castelli03}
}

\appendix
\section{Theoretical yield tables}
\label{sec:appendix}
In this section, we briefly describe the theoretical models of Type Ia supernovae.
These models are discussed in further detail in Section~4.1 of \citet{Kirby19}.
Table~\ref{tab:typeia_yields_mch} and Table~\ref{tab:typeia_yields_sub} list the theoretical [Mn/Fe] yields predicted by the $M_{\mathrm{Ch}}$ and sub-$M_{\mathrm{Ch}}$ models, respectively.
Here we discuss the details of [Mn/Fe] predictions from these models.

\subsection{Deflagration-to-detonation (DDT)}
We consider two sets of near-$M_{\mathrm{Ch}}$ deflagration-to-detonation transition (DDT) models.
Since the burning front is highly textured, we chose only multi-dimensional simulations.

\vspace{\baselineskip}
\textbf{DDT(S13):} \citet{Seitenzahl13b} (hereafter \citetalias{Seitenzahl13b}) produced 3D models of CO white dwarfs with varying numbers of off-center ignition sites, which are specified in the model names; e.g., N10 has 10 ignition sites.
More ignition sites correspond to stronger deflagration phases where $^{55}\mathrm{Mn}$ (or rather, its parent nucleus $^{55}\mathrm{Co}$) is produced, producing higher [Mn/Fe] yields.

\vspace{\baselineskip}
\textbf{DDT(L18):} \citet{Leung18} (hereafter \citetalias{Leung18}) computed 2D models with single central ignition points and a variety of central densities.
As described in Section~\ref{sec:mn}, manganese yields increase with density in near-$M_{\mathrm{Ch}}$ white dwarfs.
We also consider \citetalias{Leung18}'s model ``WDD2,'' the classic DDT model of \citet{Iwamoto99} updated with new electron capture rates.

\vspace{\baselineskip}
We note that both \citetalias{Seitenzahl13b} and \citetalias{Leung18} ran solar-metallicity and low-metallicity models.
However, their low-metallicity models do not include ``simmering,'' pre-explosion convective burning in the cores of near-$M_{\mathrm{Ch}}$ progenitors.
``Simmering'' may increase the neutron excess \citep[e.g.,][]{Piro08,Chamulak08}, effectively making the initial metallicity of a $M_\mathrm{Ch}$ Type Ia SN irrelevant below a threshold of $\sim 1/3-2/3 Z_{\odot}$ \citep{Martinez-Rodriguez16, Piro08}.
Since our most metal-rich stars are well below this threshold metallicity, when possible we interpolate the DDT models to a threshold metallicity of $\sim 1/3 Z_{\odot}$.

\subsection{Pure deflagration}

We consider two sets of pure deflagrations of near-$M_{\mathrm{Ch}}$ WDs.
These may represent Type Iax SNe \citep[e.g.,][]{Kromer15}, so their nucleosynthetic yields may not be applicable to ``normal'' Type Ia SNe.

\vspace{\baselineskip}
\textbf{def(F14):} \citet{Fink14} (hereafter \citetalias{Fink14}) produced 3D models that closely paralleled the DDT models of \citetalias{Seitenzahl13b} and varied the number of off-center sites of ignition.
As with \citetalias{Seitenzahl13b}, the number of ignition sites increases with the strength of the deflagration, increasing the [Mn/Fe] yields.

\vspace{\baselineskip}
\textbf{def(L18):} \citetalias{Leung18} computed pure deflagrations that paralleled the initial central densities as their DDT models.
As with the DDT models, manganese yields increase with density.
\citetalias{Leung18} also updated the pure-deflagration ``W7'' model of \citet{Iwamoto99}.

\subsection{Sub-$M_{\mathrm{Ch}}$}
We consider three sets of sub-$M_{\mathrm{Ch}}$ models.
Each set considers a range of sub-$M_{\mathrm{Ch}}$ WD masses, and within each set the [Mn/Fe] yield tends to decrease with increasing WD mass.
This is because, as mentioned in Section~\ref{sec:mn}, in low-mass WDs ($\lesssim1.2~M_{\odot}$) $^{55}$Co is produced at densities below nuclear statistical equilibrium.
As a result, the $^{55}$Co yield (and therefore the $^{55}$Mn yield) does not change drastically as a function of mass.
Meanwhile, the $^{56}$Ni mass does increase with mass; since $^{56}$Ni is the parent nucleus of most stable iron, the overall [Mn/Fe] ratio decreases with mass.

\vspace{\baselineskip}
\textbf{sub(L19):} \citet{Leung19} (hereafter \citetalias{Leung19}) used the same 2D code as their earlier work in \citetalias{Leung18}.
All \citetalias{Leung19} models were computed at solar metallicity, except for the 1.10~$M_{\odot}$ (``benchmark'') model, which we consider at $\mathrm{[Fe/H]}\sim-1.5$ for ease of comparison with the observationally-inferred yields.

\vspace{\baselineskip}
\textbf{sub(S18):} \citet{Shen18a} (hereafter \citetalias{Shen18a}) simulated 1D detonations of CO sub-$M_{\mathrm{Ch}}$ WDs. 
We again consider only models interpolated to a metallicity of $\mathrm{[Fe/H]}\sim-1.5$ to better compare against our observations.
They simulated C/O mass ratios of both 50/50 and 30/70, which is more physically representative of the C/O ratio in actual WDs. 

\vspace{\baselineskip}
\textbf{sub(B19):} \citet{Bravo19} (hereafter \citetalias{Bravo19}) also simulated 1D detonations starting at the centers of sub-$M_{\mathrm{Ch}}$ WDs.
They explored the effect of reducing the reaction rate of $^{12}\mathrm{C} + ^{16}\mathrm{O}$ by a factor of 10; these reduced reaction rate models are represented by $\xi_{\mathrm{CO}} = 0.9$ in Table~\ref{tab:typeia_yields_sub}, while models with the ``standard'' reaction rate have $\xi_{\mathrm{CO}} = 0.0$.

\begin{table*}
\begin{minipage}[t]{.5\linewidth}
\centering
\caption{Theoretical yields for $M_{\mathrm{Ch}}$ models.\label{tab:typeia_yields_mch}\footnote{In this table and in Table~\ref{tab:typeia_yields_sub}, models marked with asterisks (*) are ``special cases'' denoted with dashed lines in Figure~\ref{fig:theorycomparison}.}}
\begin{tabular}[t]{lcc}
\hline\hline
\\[-1.25em]
Model & $\log (Z/Z_{\sun})$ & [Mn/Fe] \\
\hline
\\[-0.75em]
\hline
\\[-1em]
\multicolumn{3}{c}{DDT(\citetalias{Seitenzahl13b})} \\
\hline
N1                                                             & \phs 0.0 & $+0.01$ \\
N3                                                             & \phs 0.0 & $-0.06$ \\
N10                                                            & \phs 0.0 & $+0.01$ \\
N100                                                           & \phs $-0.5$ & $+0.27$ \\
N200                                                           & \phs 0.0 & $+0.50$ \\
N1600                                                          & \phs 0.0 & $+0.53$ \\
\hline
\\[-1em]
\multicolumn{3}{c}{def(\citetalias{Fink14})} \\
\hline
N1def                                                          & \phs 0.0 & $+0.36$ \\
N3def                                                          & \phs 0.0 & $+0.42$ \\
N10def                                                         & \phs 0.0 & $+0.44$ \\
N100def                                                        & \phs 0.0 & $+0.48$ \\
N200def                                                        & \phs 0.0 & $+0.50$ \\
N1600def                                                       & \phs 0.0 & $+0.52$ \\
\hline
\\[-1em]
\multicolumn{3}{c}{DDT(\citetalias{Leung18})} \\
\hline
*WDD2                                                           & \phs 0.0 & $+0.15$ \\
DDT $1 \times 10^9$~g~cm$^{-3}$                                & \phs $-0.5$ & $-0.17$ \\
DDT $3 \times 10^9$~g~cm$^{-3}$                                & \phs $-0.5$ & $+0.14$ \\
DDT $5 \times 10^9$~g~cm$^{-3}$                                & \phs $-0.5$ & $+0.30$ \\
\hline
\\[-1em]
\multicolumn{3}{c}{def(\citetalias{Leung18})} \\
\hline
*W7                                                             & \phs $-0.5$ & $+0.30$ \\
def $1 \times 10^9$~g~cm$^{-3}$                                & \phs 0.0 & $+0.19$ \\
def $3 \times 10^9$~g~cm$^{-3}$                                & \phs 0.0 & $+0.39$ \\
def $5 \times 10^9$~g~cm$^{-3}$                                & \phs 0.0 & $+0.39$ \\
\hline
\end{tabular}
\end{minipage}\hfill
\begin{minipage}[t]{.5\linewidth}
\centering
\caption{Theoretical yields for sub-$M_{\mathrm{Ch}}$ models. \label{tab:typeia_yields_sub}}
\begin{tabular}[t]{lcc}
\hline\hline
\\[-1.25em]
Model & $\log (Z/Z_{\sun})$ & [Mn/Fe] \\
\hline
\\[-0.75em]
\hline
\\[-1em]
\multicolumn{3}{c}{sub(\citetalias{Leung19})} \\
\hline
$0.90~M_{\sun}$, $M_{\rm He} = 0.15~M_{\sun}$                  & \phs 0.0 & $+0.25$ \\
$0.95~M_{\sun}$, $M_{\rm He} = 0.15~M_{\sun}$                  & \phs 0.0 & $-0.13$ \\
$1.00~M_{\sun}$, $M_{\rm He} = 0.10~M_{\sun}$                  & \phs 0.0 & $-0.23$ \\
$1.05~M_{\sun}$, $M_{\rm He} = 0.10~M_{\sun}$                  & \phs 0.0 & $-0.25$ \\
*$1.10~M_{\sun}$, $M_{\rm He} = 0.10~M_{\sun}$                  &   $-1.5$ & $-0.44$ \\
$1.15~M_{\sun}$, $M_{\rm He} = 0.10~M_{\sun}$                  & \phs 0.0 & $-0.45$ \\
$1.20~M_{\sun}$, $M_{\rm He} = 0.05~M_{\sun}$                  & \phs 0.0 & $-0.34$ \\
\hline
\\[-1em]
\multicolumn{3}{c}{sub(\citetalias{Shen18a})} \\
\hline
$0.85~M_{\sun}$, ${\rm C/O} = 50/50$                           &   $-1.5$ & $-0.64$ \\
$0.90~M_{\sun}$, ${\rm C/O} = 50/50$                           &   $-1.5$ & $-0.75$ \\
$1.00~M_{\sun}$, ${\rm C/O} = 50/50$                           &   $-1.5$ & $-1.05$ \\
$1.10~M_{\sun}$, ${\rm C/O} = 50/50$                           &   $-1.5$ & $-1.33$ \\
*$0.85~M_{\sun}$, ${\rm C/O} = 30/70$                           &   $-1.5$ & $-0.55$ \\
*$0.90~M_{\sun}$, ${\rm C/O} = 30/70$                           &   $-1.5$ & $-0.73$ \\
*$1.00~M_{\sun}$, ${\rm C/O} = 30/70$                           &   $-1.5$ & $-1.00$ \\
*$1.10~M_{\sun}$, ${\rm C/O} = 30/70$                           &   $-1.5$ & $-1.26$ \\
\hline
\\[-1em]
\multicolumn{3}{c}{sub(\citetalias{Bravo19})} \\
\hline
$0.88~M_{\sun}$, $\xi_{\rm CO} = 0.9$                          &   $-1.5$ & $-0.55$ \\
$0.97~M_{\sun}$, $\xi_{\rm CO} = 0.9$                          &   $-1.5$ & $-0.81$ \\
$1.06~M_{\sun}$, $\xi_{\rm CO} = 0.9$                          &   $-1.5$ & $-1.16$ \\
$1.10~M_{\sun}$, $\xi_{\rm CO} = 0.9$                          &   $-1.5$ & $-1.28$ \\
$1.15~M_{\sun}$, $\xi_{\rm CO} = 0.9$                          &   $-1.5$ & $-1.42$ \\
*$0.88~M_{\sun}$, $\xi_{\rm CO} = 0.0$                          &   $-1.5$ & $-0.50$ \\
*$0.97~M_{\sun}$, $\xi_{\rm CO} = 0.0$                          &   $-1.5$ & $-0.81$ \\
*$1.06~M_{\sun}$, $\xi_{\rm CO} = 0.0$                          &   $-1.5$ & $-1.16$ \\
*$1.10~M_{\sun}$, $\xi_{\rm CO} = 0.0$                          &   $-1.5$ & $-1.28$ \\
*$1.15~M_{\sun}$, $\xi_{\rm CO} = 0.0$                          &   $-1.5$ & $-1.42$ \\
\hline
\end{tabular}
\end{minipage}
\end{table*}

\bibliographystyle{aasjournal}
\bibliography{mndwarfs}

\begin{thebibliography}{}
\expandafter\ifx\csname natexlab\endcsname\relax\def\natexlab#1{#1}\fi
\providecommand{\url}[1]{\href{#1}{#1}}

\bibitem[{Abohalima \& Frebel(2018)}]{Abohalima18}
Abohalima, A., \& Frebel, A. 2018, ApJS, 238, 36.
\newblock \url{doi.org/10.3847/1538-4365/aadfe9}

\bibitem[{Arnett {et~al.}(1971)Arnett, Truran, \& Woosley}]{Arnett71}
Arnett, W.~D., Truran, J.~W., \& Woosley, S.~E. 1971, ApJ, 165, 87.
\newblock \url{doi.org/10.1086/150878}

\bibitem[{Asplund {et~al.}(2009)Asplund, Grevesse, Sauval, \&
  Scott}]{Asplund09}
Asplund, M., Grevesse, N., Sauval, A.~J., \& Scott, P. 2009, ARA{\&}A, 47, 481.
\newblock \url{doi.org/10.1146/annurev.astro.46.060407.145222}

\bibitem[{Baumgardt \& Hilker(2018)}]{Baumgardt18}
Baumgardt, H., \& Hilker, M. 2018, MNRAS, 478, 1520.
\newblock \url{doi.org/10.1093/mnras/sty1057}

\bibitem[{Bergemann {et~al.}(2017)Bergemann, Collet, Sch{\"{o}}nrich, Andrae,
  Kovalev, Ruchti, Hansen, \& Magic}]{Bergemann2017}
Bergemann, M., Collet, R., Sch{\"{o}}nrich, R., {et~al.} 2017, ApJ, 847, 16.
\newblock \url{doi.org/10.3847/1538-4357/aa88b5}

\bibitem[{Bergemann \& Gehren(2008)}]{Bergemann08}
Bergemann, M., \& Gehren, T. 2008, A{\&}A, 492, 823.
\newblock \url{doi.org/10.1051/0004-6361:200810098}

\bibitem[{Bergemann {et~al.}(2012)Bergemann, Lind, Collet, Magic, \&
  Asplund}]{Bergemann2012}
Bergemann, M., Lind, K., Collet, R., Magic, Z., \& Asplund, M. 2012, MNRAS,
  427, 27.
\newblock \url{doi.org/10.1111/j.1365-2966.2012.21687.x}

\bibitem[{Bergemann {et~al.}(2019)Bergemann, Gallagher, Eitner, Bautista,
  Collet, Yakovleva, Mayriedl, Plez, Carlsson, Leenaarts, Belyaev, \&
  Hansen}]{Bergemann19}
Bergemann, M., Gallagher, A.~J., Eitner, P., {et~al.} 2019, arXiv:1905.05200.
\newblock \url{adsabs.harvard.edu/abs/arXiv:1905.05200}

\bibitem[{Bonifacio {et~al.}(2009)Bonifacio, Spite, Cayrel, Hill, Spite,
  Fran{\c{c}}ois, Plez, Ludwig, Caffau, Molaro, Depagne, Andersen, Barbuy,
  Beers, Nordstr{\"{o}}m, \& Primas}]{Bonifacio09}
Bonifacio, P., Spite, M., Cayrel, R., {et~al.} 2009, A{\&}A, 501, 519.
\newblock \url{doi.org/10.1051/0004-6361/200810610}

\bibitem[{Bravo {et~al.}(2019)Bravo, Badenes, \&
  Mart{\'{i}}nez-Rodr{\'{i}}guez}]{Bravo19}
Bravo, E., Badenes, C., \& Mart{\'{i}}nez-Rodr{\'{i}}guez, H. 2019, MNRAS, 482,
  4346.
\newblock \url{doi.org/10.1093/mnras/sty2951}

\bibitem[{Carretta {et~al.}(2010)Carretta, Bragaglia, Gratton, Lucatello,
  Bellazzini, Catanzaro, Leone, Momany, Piotto, \& D'Orazi}]{Carretta10}
Carretta, E., Bragaglia, A., Gratton, R.~G., {et~al.} 2010, A{\&}A, 520,
  arXiv:1006.5866.
\newblock \url{doi.org/10.1051/0004-6361/201014924}

\bibitem[{Castelli \& Kurucz(2003)}]{Castelli03}
Castelli, F., \& Kurucz, R.~L. 2003, in IAUS, Vol. 210 (Astronomical Society of
  the Pacific), A20.
\newblock \url{adsabs.harvard.edu/abs/2003IAUS..210P.A20C/abstract}

\bibitem[{Cescutti \& Kobayashi(2017)}]{Cescutti17}
Cescutti, G., \& Kobayashi, C. 2017, A{\&}A, 607, A23.
\newblock \url{doi.org/10.1051/0004-6361/201731398}

\bibitem[{Chamulak {et~al.}(2008)Chamulak, Brown, Timmes, \&
  Dupczak}]{Chamulak08}
Chamulak, D.~A., Brown, E.~F., Timmes, F.~X., \& Dupczak, K. 2008, ApJ, 677,
  160.
\newblock \url{doi.org/10.1086/528944}

\bibitem[{Cooper {et~al.}(2012)Cooper, Newman, Davis, Finkbeiner, \&
  Gerke}]{Cooper12}
Cooper, M.~C., Newman, J.~A., Davis, M., Finkbeiner, D.~P., \& Gerke, B.~F.
  2012, ASCL, 1203.003

\bibitem[{Cunha {et~al.}(2010)Cunha, Smith, Bergemann, Suntzeff, \&
  Lambert}]{Cunha10}
Cunha, K., Smith, V.~V., Bergemann, M., Suntzeff, N.~B., \& Lambert, D.~L.
  2010, ApJ, 717, 333.
\newblock \url{doi.org/10.1088/0004-637X/717/1/333}

\bibitem[{de~Boer {et~al.}(2012)de~Boer, Tolstoy, Hill, Saha, Olsen,
  Starkenburg, Lemasle, Irwin, \& Battaglia}]{deBoer12}
de~Boer, T. J.~L., Tolstoy, E., Hill, V., {et~al.} 2012, A{\&}A, 539, A103.
\newblock \url{doi.org/10.1051/0004-6361/201118378}

\bibitem[{Duggan {et~al.}(2018)Duggan, Kirby, Andrievsky, \&
  Korotin}]{Duggan18}
Duggan, G.~E., Kirby, E.~N., Andrievsky, S.~M., \& Korotin, S.~A. 2018, ApJ,
  869, 50.
\newblock \url{doi.org/10.3847/1538-4357/aaeb8e}

\bibitem[{Escala {et~al.}(2019)Escala, Kirby, Gilbert, Cunningham, \&
  Wojno}]{Escala19}
Escala, I., Kirby, E.~N., Gilbert, K.~M., Cunningham, E.~C., \& Wojno, J. 2019,
  ApJ, 878, 42.
\newblock \url{doi.org/10.3847/1538-4357/ab1eac}

\bibitem[{Faber {et~al.}(2003)Faber, Phillips, Kibrick, Alcott, Allen, Burrous,
  Cantrall, Clarke, Coil, Cowley, Davis, Deich, Dietsch, Gilmore, Harper,
  Hilyard, Lewis, McVeigh, Newman, Osborne, Schiavon, Stover, Tucker, Wallace,
  Wei, Wirth, \& Wright}]{Faber03}
Faber, S.~M., Phillips, A.~C., Kibrick, R.~I., {et~al.} 2003, in 2003SPIE 4841,
  ed. M.~Iye \& A.~F.~M. Moorwood, Vol. 4841, 1657.
\newblock \url{doi.org/10.1117/12.460346}

\bibitem[{Feltzing {et~al.}(2007)Feltzing, Fohlman, \& Bensby}]{Feltzing07}
Feltzing, S., Fohlman, M., \& Bensby, T. 2007, A{\&}A, 467, 665.
\newblock \url{doi.org/10.1051/0004-6361:20065797}

\bibitem[{Fink {et~al.}(2014)Fink, Kromer, Seitenzahl, Ciaraldi-Schoolmann,
  R{\"{o}}pke, Sim, Pakmor, Ruiter, \& Hillebrandt}]{Fink14}
Fink, M., Kromer, M., Seitenzahl, I.~R., {et~al.} 2014, MNRAS, 438, 1762.
\newblock \url{doi.org/10.1093/mnras/stt2315}

\bibitem[{Foreman-Mackey {et~al.}(2013)Foreman-Mackey, Hogg, Lang, \&
  Goodman}]{Foreman-Mackey13}
Foreman-Mackey, D., Hogg, D.~W., Lang, D., \& Goodman, J. 2013, PASP, 125, 306.
\newblock \url{doi.org/10.1086/670067}

\bibitem[{Frebel {et~al.}(2010)Frebel, Simon, Geha, \& Willman}]{Frebel10}
Frebel, A., Simon, J.~D., Geha, M., \& Willman, B. 2010, ApJ, 708, 560.
\newblock \url{doi.org/10.1088/0004-637X/708/1/560}

\bibitem[{Giammichele {et~al.}(2012)Giammichele, Bergeron, \&
  Dufour}]{Giammichele12}
Giammichele, N., Bergeron, P., \& Dufour, P. 2012, ApJS, 199, 29.
\newblock \url{doi.org/10.1088/0067-0049/199/2/29}

\bibitem[{Gustafsson {et~al.}(1975)Gustafsson, Bell, Eriksson, \&
  Nordlund}]{Gustafsson75}
Gustafsson, B., Bell, R.~A., Eriksson, K., \& Nordlund, A. 1975, A{\&}A, 500,
  67.
\newblock \url{adsabs.harvard.edu/abs/1975A{\%}26A....42..407G/abstract}

\bibitem[{Gustafsson {et~al.}(2008)Gustafsson, Edvardsson, Eriksson,
  J{\o}rgensen, Nordlund, \& Plez}]{Gustafsson08}
Gustafsson, B., Edvardsson, B., Eriksson, K., {et~al.} 2008, A{\&}A, 486, 951.
\newblock \url{doi.org/10.1051/0004-6361:200809724}

\bibitem[{Gustafsson {et~al.}(2003)Gustafsson, Edvardsson, Eriksson,
  Mizuno-Wiedner, Jorgensen, \& Plez}]{Gustafsson03}
Gustafsson, B., Edvardsson, B., Eriksson, K., {et~al.} 2003, in 2003ASPC 288
  (Astronomical Society of the Pacific), 697.
\newblock \url{adsabs.harvard.edu/abs/2003ASPC..288..331G/abstract}

\bibitem[{Harris(1996)}]{Harris96}
Harris, W.~E. 1996, AJ, 112, 1487.
\newblock \url{doi.org/10.1086/118116}

\bibitem[{Hernandez {et~al.}(2000)Hernandez, Gilmore, \&
  Valls-Gabaud}]{Hernandez00}
Hernandez, X., Gilmore, G., \& Valls-Gabaud, D. 2000, MNRAS, 317, 831.
\newblock \url{doi.org/10.1046/j.1365-8711.2000.03809.x}

\bibitem[{Hogg {et~al.}(2010)Hogg, Bovy, \& Lang}]{Hogg10}
Hogg, D.~W., Bovy, J., \& Lang, D. 2010, arXiv:1008.4686.
\newblock \url{adsabs.harvard.edu/abs/2010arXiv1008.4686H}

\bibitem[{Hunter(2007)}]{matplotlib}
Hunter, J.~D. 2007, Comput. Sci. Eng., 9, 90.
\newblock \url{ieeexplore.ieee.org/document/4160265/}

\bibitem[{Iben \& Tutukov(1984)}]{Iben84}
Iben, I., \& Tutukov, A.~V. 1984, ApJS, 54, 335.
\newblock \url{doi.org/10.1086/190932}

\bibitem[{Iwamoto {et~al.}(1999)Iwamoto, Brachwitz, Nomoto, Kishimoto, Umeda,
  Hix, \& Thielemann}]{Iwamoto99}
Iwamoto, K., Brachwitz, F., Nomoto, K., {et~al.} 1999, ApJS, 125, 439.
\newblock \url{doi.org/10.1086/313278}

\bibitem[{Johnson \& Pilachowski(2010)}]{Johnson10}
Johnson, C.~I., \& Pilachowski, C.~A. 2010, ApJ, 722, 1373.
\newblock \url{doi.org/10.1088/0004-637X/722/2/1373}

\bibitem[{Jones {et~al.}(2001)Jones, Oliphant, Peterson, {et~al.}}]{scipy}
Jones, E., Oliphant, T., Peterson, P., {et~al.} 2001, {SciPy: Open source
  scientific tools for Python}, , .
\newblock \url{www.scipy.org/}

\bibitem[{Khokhlov(1991)}]{Khokhlov91}
Khokhlov, A. 1991, A{\&}A, 245, 114.
\newblock \url{adsabs.harvard.edu/abs/1991A{\%}26A...245..114K}

\bibitem[{Kirby {et~al.}(2009)Kirby, Guhathakurta, Bolte, Sneden, \&
  Geha}]{Kirby09}
Kirby, E.~N., Guhathakurta, P., Bolte, M., Sneden, C., \& Geha, M.~C. 2009,
  ApJ, 705, 328.
\newblock \url{doi.org/10.1088/0004-637X/705/1/328}

\bibitem[{Kirby {et~al.}(2016)Kirby, Guhathakurta, Zhang, Hong, Guo, Guo,
  Cohen, \& Cunha}]{Kirby16}
Kirby, E.~N., Guhathakurta, P., Zhang, A.~J., {et~al.} 2016, ApJ, 819, 135.
\newblock \url{doi.org/10.3847/0004-637X/819/2/135}

\bibitem[{Kirby {et~al.}(2018)Kirby, Xie, Guo, Kovalev, \& Bergemann}]{Kirby18}
Kirby, E.~N., Xie, J.~L., Guo, R., Kovalev, M., \& Bergemann, M. 2018, ApJS,
  237, 18.
\newblock \url{doi.org/10.3847/1538-4365/aac952}

\bibitem[{Kirby {et~al.}(2010)Kirby, Guhathakurta, Simon, Geha, Rockosi,
  Sneden, Cohen, Sohn, Majewski, \& Siegel}]{Kirby10}
Kirby, E.~N., Guhathakurta, P., Simon, J.~D., {et~al.} 2010, ApJS, 191, 352.
\newblock \url{doi.org/10.1088/0067-0049/191/2/352}

\bibitem[{Kirby {et~al.}(2019)Kirby, Xie, Guo, de~los Reyes, Bergemann,
  Kovalev, Shen, Piro, \& McWilliam}]{Kirby19}
Kirby, E.~N., Xie, J.~L., Guo, R., {et~al.} 2019, ApJ, 881, 45.
\newblock \url{doi.org/10.3847/1538-4357/ab2c02}

\bibitem[{Kobayashi \& Nomoto(2009)}]{Kobayashi09}
Kobayashi, C., \& Nomoto, K. 2009, ApJ, 707, 1466.
\newblock \url{doi.org/10.1088/0004-637X/707/2/1466}

\bibitem[{Kobayashi {et~al.}(2015)Kobayashi, Nomoto, \& Hachisu}]{Kobayashi15}
Kobayashi, C., Nomoto, K., \& Hachisu, I. 2015, ApJL, 804, L24.
\newblock \url{doi.org/10.1088/2041-8205/804/1/L24}

\bibitem[{Kramida {et~al.}(2014)Kramida, Ralchenko, Reader, \& {NIST ASD
  Team}}]{Kramida14}
Kramida, A., Ralchenko, Y., Reader, J., \& {NIST ASD Team}. 2014, {NIST Atomic
  Spectra Database (ver. 5.2)}, Online,  National Institute of Standards and
  Technology.
\newblock \url{physics.nist.gov/asd}

\bibitem[{Kromer {et~al.}(2015)Kromer, Ohlmann, Pakmor, Ruiter, Hillebrandt,
  Marquardt, R{\"{o}}pke, Seitenzahl, Sim, \& Taubenberger}]{Kromer15}
Kromer, M., Ohlmann, S.~T., Pakmor, R., {et~al.} 2015, MNRAS, 450, 3045.
\newblock \url{doi.org/10.1093/mnras/stv886}

\bibitem[{Kurucz(1993)}]{Kurucz93}
Kurucz, R. 1993, ATLAS9 Stellar Atmos. Programs 2 km/s grid. Kurucz CD-ROM No.
  13. Cambridge, 13.
\newblock \url{adsabs.harvard.edu/abs/1993KurCD..13.....K}

\bibitem[{Lardo {et~al.}(2013)Lardo, Pancino, Mucciarelli, Bellazzini, Rejkuba,
  Marinoni, Cocozza, Altavilla, \& Ragaini}]{Lardo13}
Lardo, C., Pancino, E., Mucciarelli, A., {et~al.} 2013, MNRAS, 433, 1941.
\newblock \url{doi.org/10.1093/mnras/stt854}

\bibitem[{Lesaffre {et~al.}(2006)Lesaffre, Han, Tout, Podsiadlowski, \&
  Martin}]{Lesaffre06}
Lesaffre, P., Han, Z., Tout, C.~A., Podsiadlowski, P., \& Martin, R.~G. 2006,
  MNRAS, 368, 187.
\newblock \url{doi.org/10.1111/j.1365-2966.2006.10068.x}

\bibitem[{Letarte {et~al.}(2010)Letarte, Hill, Tolstoy, Jablonka, Shetrone,
  Venn, Spite, Irwin, Battaglia, Helmi, Primas, Francois, Kaufer, Szeifert,
  Arimoto, \& Sadakane}]{Letarte10}
Letarte, B., Hill, V., Tolstoy, E., {et~al.} 2010, A{\&}A, 523, A17.
\newblock \url{doi.org/10.1051/0004-6361/200913413}

\bibitem[{Leung \& Nomoto(2018)}]{Leung18}
Leung, S.-C., \& Nomoto, K. 2018, ApJ, 861, 143.
\newblock \url{doi.org/10.3847/1538-4357/aac2df}

\bibitem[{Leung \& Nomoto(2019)}]{Leung19}
---. 2019, arXiv:1901.10007.
\newblock \url{doi.org/1901.10007}

\bibitem[{Livne(1990)}]{Livne90}
Livne, E. 1990, ApJ, 354, L53.
\newblock \url{doi.org/10.1086/185721}

\bibitem[{Maoz {et~al.}(2012)Maoz, Mannucci, \& Brandt}]{Maoz12}
Maoz, D., Mannucci, F., \& Brandt, T.~D. 2012, MNRAS, 426, 3282.
\newblock \url{doi.org/10.1111/j.1365-2966.2012.21871.x}

\bibitem[{Maoz {et~al.}(2014)Maoz, Mannucci, \& Nelemans}]{Maoz14}
Maoz, D., Mannucci, F., \& Nelemans, G. 2014, ARA{\&}A, 52, 107.
\newblock \url{doi.org/10.1146/annurev-astro-082812-141031}

\bibitem[{Margutti {et~al.}(2012)Margutti, Soderberg, Chomiuk, Chevalier,
  Hurley, Milisavljevic, Foley, Hughes, Slane, Fransson, Moe, Barthelmy,
  Boynton, Briggs, Connaughton, Costa, Cummings, {Del Monte}, Enos, Fellows,
  Feroci, Fukazawa, Gehrels, Goldsten, Golovin, Hanabata, Harshman, Krimm,
  Litvak, Makishima, Marisaldi, Mitrofanov, Murakami, Ohno, Palmer, Sanin,
  Starr, Svinkin, Takahashi, Tashiro, Terada, \& Yamaoka}]{Margutti12}
Margutti, R., Soderberg, A.~M., Chomiuk, L., {et~al.} 2012, ApJ, 751, 134.
\newblock \url{doi.org/10.1088/0004-637X/751/2/134}

\bibitem[{Mart{\'{i}}nez-Rodr{\'{i}}guez
  {et~al.}(2016)Mart{\'{i}}nez-Rodr{\'{i}}guez, Piro, Schwab, \&
  Badenes}]{Martinez-Rodriguez16}
Mart{\'{i}}nez-Rodr{\'{i}}guez, H., Piro, A.~L., Schwab, J., \& Badenes, C.
  2016, ApJ, 825, 57.
\newblock \url{doi.org/10.3847/0004-637X/825/1/57}

\bibitem[{Mashonkina {et~al.}(2019)Mashonkina, Sitnova, Yakovleva, \&
  Belyaev}]{Mashonkina2019}
Mashonkina, L., Sitnova, T., Yakovleva, S.~A., \& Belyaev, A.~K. 2019, A{\&}A,
  631.
\newblock \url{doi.org/10.1051/0004-6361/201935753}

\bibitem[{Massari {et~al.}(2014)Massari, Mucciarelli, Ferraro, Origlia, Rich,
  Lanzoni, Dalessandro, Ibata, Lovisi, Bellazzini, \& Reitzel}]{Massari14}
Massari, D., Mucciarelli, A., Ferraro, F.~R., {et~al.} 2014, ApJ, 791,
  arXiv:1407.0047.
\newblock \url{doi.org/10.1088/0004-637X/791/2/101}

\bibitem[{McConnachie(2012)}]{McConnachie12}
McConnachie, A. 2012, AJ, 144, 4.
\newblock \url{doi.org/10.1088/0004-6256/144/1/4}

\bibitem[{McWilliam {et~al.}(2018)McWilliam, Piro, Badenes, \&
  Bravo}]{McWilliam18}
McWilliam, A., Piro, A.~L., Badenes, C., \& Bravo, E. 2018, ApJ, 857, 97.
\newblock \url{doi.org/10.3847/1538-4357/aab772}

\bibitem[{Newman {et~al.}(2013)Newman, Cooper, Davis, Faber, Coil,
  Guhathakurta, Koo, Phillips, Conroy, Dutton, Finkbeiner, Gerke, Rosario,
  Weiner, Willmer, Yan, Harker, Kassin, Konidaris, Lai, Madgwick, Noeske,
  Wirth, Connolly, Kaiser, Kirby, Lemaux, Lin, Lotz, Luppino, Marinoni,
  Matthews, Metevier, \& Schiavon}]{Newman13}
Newman, J.~A., Cooper, M.~C., Davis, M., {et~al.} 2013, ApJS, 208, 5.
\newblock \url{doi.org/10.1088/0067-0049/208/1/5}

\bibitem[{Nomoto(1982)}]{Nomoto82b}
Nomoto, K. 1982, ApJ, 257, 780.
\newblock \url{doi.org/10.1086/160031}

\bibitem[{North {et~al.}(2012)North, Cescutti, Jablonka, Hill, Shetrone,
  Letarte, Lemasle, Venn, Battaglia, Tolstoy, Irwin, Primas, \&
  Fran{\c{c}}ois}]{North12}
North, P., Cescutti, G., Jablonka, P., {et~al.} 2012, A{\&}A, 541, 45.
\newblock \url{doi.org/10.1051/0004-6361/201118636}

\bibitem[{Pancino {et~al.}(2011)Pancino, Mucciarelli, Sbordone, Bellazzini,
  Pasquini, Monaco, \& Ferraro}]{Pancino11}
Pancino, E., Mucciarelli, A., Sbordone, L., {et~al.} 2011, A{\&}A, 527, A18.
\newblock \url{doi.org/10.1051/0004-6361/201016024}

\bibitem[{P{\'{e}}rez-Torres {et~al.}(2014)P{\'{e}}rez-Torres, Lundqvist,
  Beswick, Bj{\"{o}}rnsson, Muxlow, Paragi, Ryder, Alberdi, Fransson, Marcaide,
  Mart{\'{i}}-Vidal, Ros, Argo, \& Guirado}]{PerezTorres14}
P{\'{e}}rez-Torres, M.~A., Lundqvist, P., Beswick, R.~J., {et~al.} 2014, ApJ,
  792, 38.
\newblock \url{doi.org/10.1088/0004-637X/792/1/38}

\bibitem[{Perlmutter {et~al.}(1999)Perlmutter, Aldering, Goldhaber, Knop,
  Nugent, Castro, Deustua, Fabbro, Goobar, Groom, Hook, Kim, Kim, Lee, Nunes,
  Pain, Pennypacker, Quimby, Lidman, Ellis, Irwin, McMahon, Ruiz‐Lapuente,
  Walton, Schaefer, Boyle, Filippenko, Matheson, Fruchter, Panagia, Newberg,
  Couch, \& Project}]{Perlmutter99}
Perlmutter, S., Aldering, G., Goldhaber, G., {et~al.} 1999, ApJ, 517, 565.
\newblock \url{doi.org/10.1086/307221}

\bibitem[{Phillips(1993)}]{Phillips93}
Phillips, M.~M. 1993, ApJ, 413, L105.
\newblock \url{adsabs.harvard.edu/doi/10.1086/186970}

\bibitem[{Piro \& Bildsten(2008)}]{Piro08}
Piro, A.~L., \& Bildsten, L. 2008, ApJ, 673, 1009.
\newblock \url{doi.org/10.1086/524189}

\bibitem[{Riess {et~al.}(1998)Riess, Filippenko, Challis, Clocchiattia,
  Diercks, Garnavich, Gilliland, Hogan, Jha, Kirshner, Leibundgut, Phillips,
  Reiss, Schmidt, Schommer, Smith, Spyromilio, Stubbs, Suntzeff, \&
  Tonry}]{Riess98}
Riess, A.~G., Filippenko, A.~V., Challis, P., {et~al.} 1998, AJ, 116, 1009.
\newblock \url{doi.org/10.1086/300499}

\bibitem[{Robitaille {et~al.}(2013)Robitaille, Tollerud, Greenfield,
  Droettboom, Bray, Aldcroft, Davis, Ginsburg, Price-Whelan, Kerzendorf,
  Conley, Crighton, Barbary, Muna, Ferguson, Grollier, Parikh, Nair,
  G{\"{u}}nther, Deil, Woillez, Conseil, Kramer, Turner, Singer, Fox, Weaver,
  Zabalza, Edwards, {Azalee Bostroem}, Burke, Casey, Crawford, Dencheva, Ely,
  Jenness, Labrie, Lim, Pierfederici, Pontzen, Ptak, Refsdal, Servillat, \&
  Streicher}]{astropy}
Robitaille, T.~P., Tollerud, E.~J., Greenfield, P., {et~al.} 2013, A{\&}A, 558,
  A33.
\newblock \url{doi.org/10.1051/0004-6361/201322068}

\bibitem[{Ruiter {et~al.}(2011)Ruiter, Belczynski, Sim, Hillebrandt, Fryer,
  Fink, \& Kromer}]{Ruiter11}
Ruiter, A.~J., Belczynski, K., Sim, S.~A., {et~al.} 2011, MNRAS, 417, 408.
\newblock \url{doi.org/10.1111/j.1365-2966.2011.19276.x}

\bibitem[{Seitenzahl {et~al.}(2013{\natexlab{a}})Seitenzahl, Cescutti,
  R{\"{o}}pke, Ruiter, \& Pakmor}]{Seitenzahl13}
Seitenzahl, I.~R., Cescutti, G., R{\"{o}}pke, F.~K., Ruiter, A.~J., \& Pakmor,
  R. 2013{\natexlab{a}}, A{\&}A, 559, L5.
\newblock \url{doi.org/10.1051/0004-6361/201322599}

\bibitem[{Seitenzahl {et~al.}(2009)Seitenzahl, Taubenberger, \&
  Sim}]{Seitenzahl09}
Seitenzahl, I.~R., Taubenberger, S., \& Sim, S.~A. 2009, MNRAS, 400, 531.
\newblock \url{doi.org/10.1111/j.1365-2966.2009.15478.x}

\bibitem[{Seitenzahl \& Townsley(2017)}]{Seitenzahl17}
Seitenzahl, I.~R., \& Townsley, D.~M. 2017, in Handb. Supernovae, ed.
  A.~Alsabti \& P.~Murdin (Springer, Cham), 1955.
\newblock \url{doi.org/10.1007/978-3-319-20794-0{\_}87-2}

\bibitem[{Seitenzahl {et~al.}(2013{\natexlab{b}})Seitenzahl,
  Ciaraldi-Schoolmann, R{\"{o}}pke, Fink, Hillebrandt, Kromer, Pakmor, Ruiter,
  Sim, \& Taubenberger}]{Seitenzahl13b}
Seitenzahl, I.~R., Ciaraldi-Schoolmann, F., R{\"{o}}pke, F.~K., {et~al.}
  2013{\natexlab{b}}, MNRAS, 429, 1156.
\newblock \url{doi.org/10.1093/mnras/sts402}

\bibitem[{Seitenzahl {et~al.}(2015)Seitenzahl, Summa, Krau{\ss}, Sim, Diehl,
  Els{\"{a}}sser, Fink, Hillebrandt, Kromer, Maeda, Mannheim, Pakmor,
  R{\"{o}}pke, Ruiter, \& Wilms}]{Seitenzahl15}
Seitenzahl, I.~R., Summa, A., Krau{\ss}, F., {et~al.} 2015, MNRAS, 447, 1484.
\newblock \url{doi.org/10.1093/mnras/stu2537}

\bibitem[{Seitenzahl {et~al.}(2016)Seitenzahl, Kromer, Ohlmann,
  Ciaraldi-Schoolmann, Marquardt, Fink, Hillebrandt, Pakmor, Roepke, Ruiter,
  Sim, \& Taubenberger}]{Seitenzahl16}
Seitenzahl, I.~R., Kromer, M., Ohlmann, S.~T., {et~al.} 2016, A{\&}A, 592, A57.
\newblock \url{doi.org/10.1051/0004-6361/201527251}

\bibitem[{Shen \& Bildsten(2007)}]{Shen07}
Shen, K.~J., \& Bildsten, L. 2007, ApJ, 660, 1444.
\newblock \url{doi.org/10.1086/513457}

\bibitem[{Shen {et~al.}(2018{\natexlab{a}})Shen, Kasen, Miles, \&
  Townsley}]{Shen18a}
Shen, K.~J., Kasen, D., Miles, B.~J., \& Townsley, D.~M. 2018{\natexlab{a}},
  AJ, 854, 52.
\newblock \url{doi.org/10.3847/1538-4357/aaa8de}

\bibitem[{Shen {et~al.}(2018{\natexlab{b}})Shen, Boubert, G{\"{a}}nsicke, Jha,
  Andrews, Chomiuk, Foley, Fraser, Gromadzki, Guillochon, Kotze, Maguire,
  Siebert, Smith, Strader, Badenes, Kerzendorf, Koester, Kromer, Miles, Pakmor,
  Schwab, Toloza, Toonen, Townsley, \& Williams}]{Shen18b}
Shen, K.~J., Boubert, D., G{\"{a}}nsicke, B.~T., {et~al.} 2018{\natexlab{b}},
  ApJ, 865, 15.
\newblock \url{doi.org/10.3847/1538-4357/aad55b}

\bibitem[{Shetrone {et~al.}(2003)Shetrone, Venn, Tolstoy, Primas, Hill, \&
  Kaufer}]{Shetrone03}
Shetrone, M., Venn, K.~A., Tolstoy, E., {et~al.} 2003, AJ, 125, 684.
\newblock \url{doi.org/10.1086/345966}

\bibitem[{Simon \& Geha(2007)}]{Simon07}
Simon, J.~D., \& Geha, M. 2007, ApJ, 670, 313.
\newblock \url{doi.org/10.1086/521816}

\bibitem[{Sneden {et~al.}(2012)Sneden, Bean, Ivans, Lucatello, \&
  Sobeck}]{moog}
Sneden, C., Bean, J., Ivans, I., Lucatello, S., \& Sobeck, J. 2012, Astrophys.
  Source Code Libr., 1202.009.
\newblock \url{adsabs.harvard.edu/abs/2012ascl.soft02009S}

\bibitem[{Sneden {et~al.}(2016)Sneden, Cowan, Kobayashi, Pignatari, Lawler,
  {Den Hartog}, \& Wood}]{Sneden16}
Sneden, C., Cowan, J.~J., Kobayashi, C., {et~al.} 2016, ApJ, 817, 53.
\newblock \url{doi.org/10.3847/0004-637X/817/1/53}

\bibitem[{Sneden {et~al.}(1997)Sneden, Kraft, Shetrone, Smith, Langer, \&
  Prosser}]{Sneden97}
Sneden, C., Kraft, R.~P., Shetrone, M.~D., {et~al.} 1997, AJ, 114, 1964.
\newblock \url{doi.org/10.1086/118618}

\bibitem[{Sobeck {et~al.}(2006)Sobeck, Ivans, Simmerer, Sneden, Hoeflich,
  Fulbright, \& Kraft}]{Sobeck06}
Sobeck, J.~S., Ivans, I.~I., Simmerer, J.~A., {et~al.} 2006, AJ, 131, 2949.
\newblock \url{doi.org/10.1086/503106}

\bibitem[{Spite(1967)}]{Spite67}
Spite, M. 1967, Ann. d'Astrophysique, 30, 211.
\newblock \url{adsabs.harvard.edu/abs/1967AnAp...30..211S/abstract}

\bibitem[{Tremblay {et~al.}(2016)Tremblay, Cummings, Kalirai, G{\"{a}}nsicke,
  Gentile-Fusillo, \& Raddi}]{Tremblay16}
Tremblay, P.-E., Cummings, J., Kalirai, J.~S., {et~al.} 2016, MNRAS, 461, 2100.
\newblock \url{doi.org/10.1093/mnras/stw1447}

\bibitem[{Webbink(1984)}]{Webbink84}
Webbink, R.~F. 1984, ApJ, 277, 355.
\newblock \url{doi.org/10.1086/161701}

\bibitem[{Weisz {et~al.}(2014)Weisz, Dolphin, Skillman, Holtzman, Gilbert,
  Dalcanton, \& Williams}]{Weisz14}
Weisz, D.~R., Dolphin, A.~E., Skillman, E.~D., {et~al.} 2014, ApJ, 789, 147.
\newblock \url{doi.org/10.1088/0004-637X/789/2/147}

\bibitem[{Woosley \& Kasen(2011)}]{Woosley11}
Woosley, S.~E., \& Kasen, D. 2011, ApJ, 734, 38.
\newblock \url{doi.org/10.1088/0004-637X/734/1/38}

\bibitem[{Woosley {et~al.}(1986)Woosley, Taam, \& Weaver}]{Woosley86}
Woosley, S.~E., Taam, R.~E., \& Weaver, T.~A. 1986, ApJ, 301, 601.
\newblock \url{doi.org/10.1086/163926}

\bibitem[{Yamaguchi {et~al.}(2015)Yamaguchi, Badenes, Foster, Bravo, Williams,
  Maeda, Nobukawa, Eriksen, Brickhouse, Petre, \& Koyama}]{Yamaguchi15}
Yamaguchi, H., Badenes, C., Foster, A.~R., {et~al.} 2015, ApJ, 801, L31.
\newblock \url{doi.org/10.1088/2041-8205/801/2/L31}

\bibitem[{Yong {et~al.}(2014)Yong, Roederer, Grundahl, {Da Costa}, Karakas,
  Norris, Aoki, Fishlock, Marino, Milone, \& Shingles}]{Yong14}
Yong, D., Roederer, I.~U., Grundahl, F., {et~al.} 2014, MNRAS, 441, 3396.
\newblock \url{doi.org/10.1093/mnras/stu806}

\end{thebibliography}

\end{document}